\documentclass[12pt, a4paper]{article}
\usepackage[utf8]{inputenc}
\usepackage[T1]{fontenc}
\usepackage{geometry}
\usepackage{amsmath, amssymb, amsthm}
\usepackage{graphicx}
\usepackage{booktabs} 
\usepackage{hyperref} 
\usepackage{natbib}   
\usepackage{caption}
\usepackage{float}    
\usepackage{enumitem} 
\usepackage{longtable} 
\usepackage{array}
\usepackage{setspace}
\usepackage{booktabs}
\usepackage{threeparttable}
\usepackage{tabularx}  
\newcolumntype{L}{>{\raggedright\arraybackslash}X}
\newcolumntype{C}{>{\centering\arraybackslash}X}
\newcolumntype{R}{>{\raggedleft\arraybackslash}X}
\newcolumntype{Y}{>{\centering\arraybackslash}X}
\newtheorem{proposition}{Proposition}

\hypersetup{colorlinks=true, urlcolor=blue, linkcolor=blue, citecolor=blue}

\title{\textbf{A Blessing in Disguise? DeFi Exploits and Short-Horizon Responses in U.S. Commercial Paper Spreads}\thanks{
We are deeply grateful to Professor Andrei Shleifer (Harvard University) for his invaluable guidance and constructive advice.  We also thank him for his strategic advice on journal submission and future research directions.

\par We extend our gratitude to Professor Wanlin Lin (HKU-ICCDS) and Ruoran Lai (Sun Yat-sen University) for their stimulating academic discussions and insightful feedback, which significantly benefited the development of our theoretical arguments.

\par We also acknowledge the valuable assistance of Karl Yu (Boston College) and Huanxi Zhang (University of Wisconsin-Madison) for their help in refining the writing and revising the manuscript.

\par Special thanks go to Dr. Ke Dong and the faculty members at the PBC School of Finance, Tsinghua University (Tsinghua PBCSF) for their time and effort in conducting a preliminary review of the manuscript and providing helpful comments. All remaining errors are our own.
}}

\author{
    Tingyi Lin\thanks{Central University of Finance and Economics. Email: \href{mailto:hayashi@email.cufe.edu.cn}{hayashi@email.cufe.edu.cn}}
}

\date{}

\begin{document}

\maketitle
\begin{center}
\textit{First version: June, 2024}\\
\textit{This version: December,2025}\\
\href{https://ssrn.com/abstract=5935576}{Click here for the latest version}
\end{center}
\begin{abstract}
Do vulnerabilities in Decentralized Finance (DeFi) destabilize traditional short-term funding markets? While the prevailing ``Contagion Hypothesis'' posits that stablecoin reserve liquidations may transmit distress to traditional markets through fire-sale pressure, we document a short-horizon ``Flight-to-Quality'' pattern in the opposite direction. In the wake of major DeFi exploits, spreads on 3-month AA-rated commercial paper (CP) tend to narrow rather than widen. We interpret this pattern as consistent with a ``liquidity-recycling'' channel: capital leaving DeFi may be re-intermediated into traditional cash-management markets, with regulatory segmentation under SEC Rule 2a-7 making prime-eligible paper a plausible marginal destination. Because we do not directly observe daily fund-level routing into prime money market funds, this mechanism is inferred from pricing patterns and monthly holdings evidence rather than directly identified. The result is specific to exploit-driven operational shocks, this U.S. CP spread, and short event windows.

\vspace{1em}
\end{abstract}

\newpage
\begin{quote}
    \itshape
    ``All discord, harmony not understood; All partial evil, universal good.''
    \par
    \hfill --- Alexander Pope, \textit{An Essay on Man} (1733)
\end{quote}
\doublespacing
\section{Introduction}
The growing interface between Decentralized Finance (DeFi) and traditional
financial markets has intensified regulatory concern about potential spillovers from
crypto-native distress into core funding markets. A common concern is a
``contagion'' scenario: if DeFi turmoil triggers stablecoin redemptions, reserve
liquidation could increase the supply of short-term safe assets and place upward
pressure on money-market spreads. This concern is especially salient for
fiat-backed stablecoins whose reserves are invested in traditional safe and near-safe
instruments.

This paper studies a narrower question. Using the top 50 exploit-driven DeFi
operational shocks between 2021 and 2024, we ask whether such events are
associated with short-horizon movements in one specific traditional funding-market
price: the spread on 3-month AA-rated nonfinancial commercial paper (CP) relative
to the 3-month Treasury bill. The main empirical pattern is that this spread tends
to narrow on the event day and, more weakly, on the following trading day.

We organize the evidence in three layers. First, because the dependent variable is a
single aggregate daily spread, we treat stacked event-time estimates primarily as a
way to summarize the dynamic pattern around exploits. To address the pseudo-panel
concern directly, we bring forward an existing time-series local-projection
specification \citep{jorda2005estimation} estimated on the single daily spread series with
Newey-West inference. Second, we use a granular-IV design, already implemented in
the manuscript, to ask whether larger exploit-linked redemption pressure is
associated with larger spread responses. Third, we examine monthly money-market-fund
holdings to assess whether the pricing pattern is consistent with prime-fund
segmentation under Rule 2a-7.

Our preferred interpretation is a flight-to-quality or ``liquidity-recycling''
channel. Under this interpretation, exploit shocks increase demand for short-term
safe or prime-eligible instruments quickly enough that, in the specific AA CP
segment we study, buying pressure can outweigh direct reserve-sale pressure. But an
important limitation is that we do not directly observe daily fund-level routing
into prime MMFs, so the prime-MMF channel is inferred from pricing patterns and
monthly holdings evidence rather than directly identified at the daily frequency
relevant for the event-window results.

The paper also develops a stylized ambiguity-based portfolio model and a stylized
global-game run framework. These theoretical appendices are used to motivate the
possibility of demand amplification and state dependence; they are not presented as
direct structural estimators of the empirical coefficients. In particular, the
parameter $\eta$ should be interpreted as a demand-amplification parameter or panic
multiplier, not as a directly estimated structural parameter.

The contribution is therefore narrower than overturning the broad DeFi-contagion
debate. The paper documents a short-horizon pricing pattern in one U.S.
money-market spread during exploit-driven operational shocks, shows that existing
time-series and IV evidence point in the same direction, and argues that a
prime-eligible flight-to-quality interpretation is plausible. The findings should
not be extrapolated mechanically to solvency crises such as TerraUSD or FTX, where
the relevant transmission forces may be very different.

\section{Institutional Background and Stylized Facts}

\subsection{The DeFi-TradFi Nexus: Stablecoins as the Conduit}

While Decentralized Finance (DeFi) is built upon blockchain technology, its value proposition does not exist in a vacuum. Stablecoins (e.g., USDC and USDT) constitute the critical nexus connecting the DeFi ecosystem with U.S. short-term funding markets. Unlike their algorithmic variants which have shown inherent instability \citep{clements2021built,krause2025algorithmic}, fiat-collateralized stablecoins maintain a 1:1 peg to the U.S. dollar by holding substantial high-liquidity traditional financial instruments.

The literature widely characterizes stablecoins as a novel form of shadow banking \citep{gorton2021taming}. Recent studies have further scrutinized the mechanisms maintaining-or breaking-this peg, including the role of safe-asset prices, devaluation risk, public-information sensitivity, and arbitrage centralization \citep{lyons2023what,ahmed2025stablecoins,eichengreen2023stablecoin,lee2024stablecoin,ahmed2024public,ma2025stablecoin}.

Further emphasizing the TradFi link, \citep{liao2022stablecoins} discuss how stablecoin adoption can interact with traditional funding intermediation, with effects that depend on reserve composition. Public reserve disclosures and the policy literature indicate material holdings of short-term dollar instruments, though the mix varies across issuers and over time. This makes spillovers from large on-chain DeFi outflows into off-chain cash-management markets plausible, but it does not imply a mechanical one-for-one pass-through into any single money-market instrument. To identify the specific conduit of this transmission, we must account for the strict regulatory segmentation of the MMF industry.

(1) Under SEC Rule 2a‑7, a government money market fund is defined as one that invests at least 99.5\% of total assets in cash, government securities, and/or repurchase agreements that are fully collateralized (by cash or government securities). 
(2) By contrast, prime money market funds primarily invest in taxable short‑term corporate and bank debt instruments, such as commercial paper (CP) and certificates of deposit (CDs). 
(3) The MMF holdings evidence we use comes from the regulatory holdings reporting infrastructure (e.g., Form N‑MFP and related aggregates), including the Federal Reserve’s EFA holdings detail and SEC‑reported MMF datasets; this allows our subsequent identification strategies to rely on observed holdings rather than institutional assumptions alone.

The institutional facts above motivate why prime MMFs are a plausible marginal conduit for any demand response that loads on high-quality short-term private paper. They do not, by themselves, establish that exploit-related DeFi outflows are routed into prime MMFs at the daily frequency relevant for our event-window results. That latter step remains an interpretation supported indirectly by pricing patterns and monthly holdings evidence.

\subsection{Stylized Facts}
\begin{figure}[H]
    \centering
    \includegraphics[width=\textwidth, keepaspectratio]{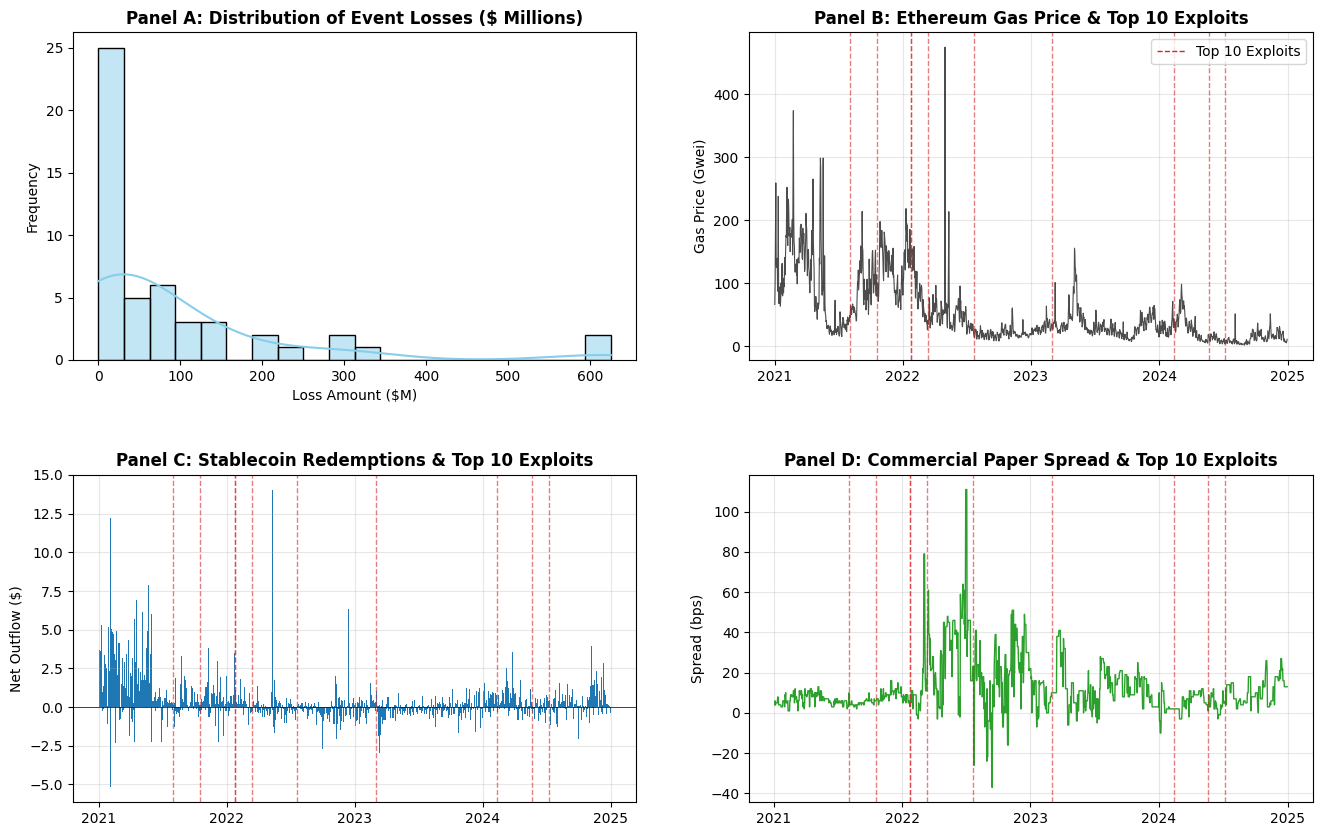} 
    \caption{\textbf{Distribution of Event Losses and Market Response.} \\ \small Panel A: Fat-tailed Distribution of Event Losses. Panel B: Instantaneous Response of Payment Friction (Gas Price). Panel C: Stablecoin Redemptions and Top 10 Exploits. Panel D: Anomalous Spread Narrowing and ``Blessing in Disguise''. The red dashed lines mark the Top 10 exploits.}
    \label{fig:stylized_facts}
\end{figure}

To visualize the dynamic linkage between DeFi operational risks and macro-financial indicators, we plot data from 2021 to 2024. Figure \ref{fig:stylized_facts} illustrates the distributional characteristics of DeFi security incidents and their temporal relationship with key financial variables.

Figure 1 comprises four panels, each depicting a distinct segment of the risk transmission chain. The red dashed lines mark the occurrence of the Top 10 exploits by loss magnitude.
\begin{itemize}
    \item \textbf{Panel A:} Fat-tailed Distribution of Event Losses. The histogram reveals a distinct long-tail distribution: the vast majority of security incidents result in relatively minor losses. Conversely, the right tail exhibits extreme outliers, with rare ``nuclear-level'' events exceeding \$200 million.
    \item \textbf{Panel B:} Instantaneous Response of Payment Friction. Gas fees frequently exhibit instantaneous, sharp spikes coinciding with major attacks, indicating physical network congestion.
    \item \textbf{Panel C:} Redemptions and Capital Flight. Massive withdrawal of capital from DeFi protocols (USDC + USDT redemptions) follows major hacks, constituting a ``digital run.''
    \item \textbf{Panel D:} Anomalous Spread Narrowing. In the short window following major attacks, the CP Spread exhibits a counter-intuitive downward trend (narrowing), contrary to contagion theory.
\end{itemize}

\begin{table}[H]
\centering
\caption{Summary Statistics}
\label{tab:summary_stats}
\begin{tabularx}{\textwidth}{L C C C C C C C}
\toprule
\textbf{Variable} & \textbf{N} & \textbf{Mean} & \textbf{Std. Dev.} & \textbf{Min} & \textbf{P25} & \textbf{Median} & \textbf{Max} \\
\midrule
\multicolumn{8}{l}{\textit{Panel A: Characteristics of DeFi Security Incidents ($N=50$)}} \\
Loss Amount (\$ Mil) & 50 & 88.80 & 136.44 & 0.24 & 5.35 & 40.50 & 625.00 \\
Log\_Loss & 50 & 16.98 & 1.97 & 12.39 & 15.49 & 17.49 & 20.25 \\
Gas Price (Gwei) & 50 & 68.44 & 59.05 & 9.19 & 19.63 & 42.01 & 215.00 \\
\midrule
\multicolumn{8}{l}{\textit{Panel B: Daily Market Data (2021-2024)}} \\
CP\_Spread (bps) & 1,005 & 12.34 & 13.32 & -26.00 & 5.00 & 8.00 & 111.00 \\
VIX Index & 1,005 & 19.44 & 5.28 & 11.86 & 15.37 & 18.51 & 38.57 \\
BTC Return (\%) & 1,004 & 0.18 & 3.94 & -22.68 & -1.81 & 0.04 & 21.11 \\
DXY Index & 1,005 & 101.04 & 5.80 & 89.44 & 96.03 & 102.91 & 114.11 \\
Net Redemption (\$ Mil) & 1,005 & -32.31 & 103.24 & -2040.67 & -77.96 & 0.00 & 5402.17 \\
\bottomrule
\end{tabularx}
\end{table}

Table I reports the summary statistics for the key variables employed in our empirical analysis based on real market data. This summary aims to illustrate the distributional characteristics of the data, specifically highlighting the extreme nature of DeFi risks and the volatility of traditional financial markets, thereby laying the empirical groundwork for our subsequent regression analysis.
\begin{table}[H]
\centering
\caption{Correlation Matrix of Key Variables}
\label{tab:correlation}
\begin{tabularx}{\textwidth}{L C C C C C}
\toprule
\textbf{Variables} & \textbf{CP\_Spread} & \textbf{Gas\_Price} & \textbf{VIX} & \textbf{BTC\_Ret} & \textbf{DXY} \\
\midrule
CP\_Spread & 1.000 & & & & \\
Gas\_Price & $-0.138^*$ & 1.000 & & & \\
VIX & $0.433^{***}$ & $0.171^{***}$ & 1.000 & & \\
BTC\_Ret & $-0.055$ & $-0.145^{***}$ & $-0.119^{***}$ & 1.000 & \\
DXY & $0.302^{***}$ & $-0.614^{***}$ & $0.103^{**}$ & 0.032 & 1.000 \\
\bottomrule
\end{tabularx}
\footnotesize \\ \textit{Notes:} Pairwise correlations using real daily data. Significance levels: *** $p < 0.01$, ** $p < 0.05$, * $p < 0.1$.
\end{table}

Table II reveals a contrast: while general market panic (VIX) is positively correlated with spreads, event-day gas prices are negatively correlated with spreads in the raw data. This pattern is suggestive of a distinct DeFi-linked flight-to-quality channel, but the correlation itself is descriptive and should not be read as a stand-alone identification result.

\section{A Stylized Framework for DeFi Redemptions and Short-Term Funding Spreads}

This section presents a stylized framework that organizes the paper's interpretation of the data. The objective is not to identify structural parameters directly from the empirical specifications, but to clarify three ideas: (i) exploit shocks can raise transaction frictions on-chain, (ii) redemptions may become state dependent when panic motives interact with congestion costs, and (iii) in a segmented short-term funding market, demand for prime-eligible paper may move differently from the broader reserve basket. The framework is therefore interpretive and deliberately narrower than the empirical reduced-form evidence.

\subsection{Sector I: DeFi Settlement Layer and Payment Frictions (The Friction Channel)}
Drawing on DeFi ``composability risk'' \citep{werner2021sok}, we model hacks as exogenous shocks to the settlement capacity of the blockchain network.

At time $t$, let the attack intensity be denoted by:
\begin{equation}
    I_t \in [0, \infty)
\end{equation}
This represents the scale of financial loss caused by external hacks or internal code vulnerabilities.

Following the analysis of single points of failure in cross-chain bridges \citep{belchior2021survey}, attack intensity directly impairs the network's ``Bridge Availability,'' denoted as:
\begin{equation}
    \Omega_t = 1 - \kappa \cdot I_t
\end{equation}
where $\kappa > 0$ is a system vulnerability parameter. A decline in network availability reflects compromised throughput or a security halt.

Furthermore, consistent with Ethereum's congestion pricing mechanism (EIP-1559), a decline in availability triggers a Priority Gas Auction (PGA) for block space (analogous to paying higher fees to withdraw funds during a bank run). This pushes up the payment friction cost, $\Phi_t$ (i.e., Gas fees). We model this relationship as a non-linear function:
\begin{equation}
    \Phi_t = \phi_0 + \phi_1 (1 - \Omega_t)^\gamma
\end{equation}
Where:
\begin{itemize}
    \item $\phi_0$ is the baseline transaction cost;
    \item $\phi_1 > 0$ is the sensitivity coefficient;
    \item $\gamma \ge 1$ captures the non-linear escalation characteristics of congestion costs.
\end{itemize}

\begin{description}
    \item[Proposition 1:] \textit{Attack intensity non-linearly drives up payment frictions by impairing network availability.}
\end{description}

\subsection{Sector II: The Stablecoin Sector: Reserves and Runs}
This section constructs a micro-model to describe the redemption behavior of DeFi investors. We adopt the view of stablecoins as shadow banking debt \citep{gorton2021taming}, combine the run framework of \citep{gorton2012securitized} with the technical friction theory of \citep{werner2021sok}, and draw on the information-sensitivity logic of \citep{gorton2014collateral}. Systemic stablecoins can trigger fire sales when redemptions spike \citep{gross2025par}. Similarly, lessons from the collapse of algorithmic stablecoins like TerraUSD suggest that once a critical threshold of confidence is breached, the ``death spiral'' is often accelerated by endogenous feedback loops \citep{krause2025algorithmic}.

\subsubsection{Normalized redemption demand and realized redemptions}

To motivate aggregate redemption pressure, we use a normalized representative-investor framework. The object of interest is not a literal binary choice by a single investor, but an aggregate desired-redemption intensity. Let $R_t^d$ denote normalized desired redemptions following exploit shock $I_t$.

We model desired redemptions as
\begin{equation}
    R_t^d = \rho_0 + \rho_1 (-\Delta p_t^{crypto}) + \rho_2 \cdot \mathbb{I}\{\Omega_t < \bar{\Omega}\},
\end{equation}
where $\Delta p_t^{crypto}$ is the crypto return and $\mathbb{I}\{\Omega_t < \bar{\Omega}\}$ is a reduced-form state indicator capturing the idea that exploit shocks can push the system into a more information-sensitive regime. The coefficient $\rho_2 > 0$ is a panic term. In Appendix B, we show in a stylized robust-control model why ambiguity can generate a discrete increase in the desired exit motive, but we do not claim that $\rho_2$ is structurally estimated in the data.

Actual realized redemptions are constrained by congestion costs:
\begin{equation}
    R_t = \frac{R_t^d}{1 + \psi \Phi_t},
\end{equation}
where $\Phi_t$ is the payment-friction measure from Sector I and $\psi > 0$ measures sensitivity to those costs.

This setup allows exploit shocks to generate state-dependent realized redemptions. When congestion is elevated but panic is limited, the denominator effect can dominate and observed redemptions are dampened. When panic becomes sufficiently strong, the numerator effect can dominate and realized redemptions may remain elevated despite high congestion.

\begin{description}
    \item[Proposition 2 (State-dependent redemption pressure):] \textit{On-chain frictions can dampen realized redemptions in ordinary exploit episodes, but this damping effect need not survive when exploit shocks push investors into a more information-sensitive panic state.}
\end{description}

\subsection{Sector III: U.S. Short-Term Funding Markets: Relative-Price Channels}

\subsubsection{Market structure and relative-price mapping}

The empirical object in this paper is not the price of a generic ``safe-asset basket,'' but the spread between 3-month AA nonfinancial commercial paper and the 3-month Treasury bill. We therefore model the relevant short-run market as a segmented relative-price problem between two instruments:

\begin{itemize}
    \item $H$: prime-eligible high-grade private paper, proxied empirically by 3-month AA nonfinancial CP;
    \item $T$: government paper, proxied by the 3-month Treasury bill.
\end{itemize}

Let $R_t$ denote gross stablecoin redemptions. A portion of reserve liquidation loads on the $H$ segment:
\begin{equation}
    \Delta S_t^{H} = s R_t, \qquad s \in [0,1].
\end{equation}
Here $s$ captures how much of issuer reserve sales mechanically translates into effective supply pressure in the AA CP segment we study.

Let exploit-induced flight-to-quality demand into the $H$ segment be
\begin{equation}
    \Delta D_t^{H} = \eta R_t,
\end{equation}
where $\eta$ is a \emph{demand-amplification parameter} (or panic multiplier), not a price elasticity. It measures the gross demand response for the $H$ segment generated per unit of redemption pressure.

The net quantity imbalance in the $H$ segment is therefore
\begin{equation}
    \Delta B_{t}^{H,\text{net}} = \Delta S_t^{H} - \Delta D_t^{H} = (s-\eta)R_t.
\end{equation}

In a segmented short-term funding market, the CP--Tbill spread responds to this relative imbalance:
\begin{equation}
    \Delta Spread_t = \lambda_H (s-\eta)R_t, \qquad \lambda_H > 0,
\end{equation}
where $\lambda_H$ is a reduced-form price-impact parameter for the AA CP segment.

Under the normalization $s=1$, this reduces to
\begin{equation}
    \Delta Spread_t = \lambda_H (1-\eta)R_t.
\end{equation}

\subsubsection{Interpretation}

This mapping clarifies two points that are important for the empirical sections. First, the paper studies the relative price of AA CP versus T-bills, not the entire reserve basket. Second, $\eta R_t$ is the \emph{gross} flight-to-quality demand response, whereas net excess demand in the normalized case is $(\eta-1)R_t$.

\begin{description}
    \item[Proposition 3 (Spread narrowing condition):] \textit{In the segmented market for prime-eligible high-grade private paper, spread narrowing occurs whenever exploit-induced demand amplification into that segment exceeds the effective supply pressure loading on the same segment, i.e. whenever $\eta > s$ (or $\eta > 1$ under the normalization $s=1$).}
\end{description}

\paragraph{Connection to the empirical analysis.}
The empirical sections do not directly estimate $\eta$ or $s$. Instead, they ask whether the observed spread response is consistent with a configuration in which exploit-linked demand for the AA CP segment dominates exploit-linked supply pressure in that same segment.

\section{Data and Identification Strategy}

\subsection{Data Sources and Sample Construction}
To study exploit-driven operational shocks and short-horizon responses in U.S. short-term funding markets, we construct a matched dataset integrating high-frequency event data, on-chain capital flows, and traditional money-market indicators. The sample spans January 1, 2021, to December 31, 2024.

\textbf{Core explanatory variable: DeFi security incidents.} Data on hacks and protocol vulnerabilities are sourced from the DeFiLlama Hacks Dashboard and the Rekt Database. These databases report exploit timestamps, names of compromised protocols, affected blockchains, and loss amounts in USD. We rank events by loss amount and select the top 50 major security incidents as the core sample.

\textbf{Event dating.} The baseline event list uses the first market-relevant date for U.S. investors. For most exploits this is the occurrence date. For disclosure-lag cases such as Ronin, it is the first public disclosure date. When the relevant timestamp falls after U.S. market hours or on a non-trading day, the event is aligned to the next U.S. trading day.

\textbf{Dependent variable: AA CP spread.} The core outcome is the spread between the ``3-Month AA Nonfinancial Commercial Paper Rate'' and the ``3-Month Treasury Bill Yield,'' sourced from the Federal Reserve Bank of St. Louis (FRED). This measure is the paper's single aggregate daily price series and should be interpreted narrowly as short-term funding conditions in the specific AA CP segment studied here.

To isolate the influence of macro-market volatility, we obtain the CBOE Volatility Index (VIX), TED spread, the U.S. Dollar Index (DXY), S\&P500 Returns, and Bitcoin Returns from Yahoo Finance as control variables. All time-series data are aligned to U.S. trading days. For the core CP-spread outcome, we rely on realized trading-day observations rather than creating synthetic spread values for non-trading days. When event timestamps fall outside U.S. market hours or on weekends, they are mapped to the next relevant U.S. trading day; auxiliary controls are aligned to that same trading-day index.

\subsection{Econometric Specification}
Given the temporal discreteness of exploit events and the high-frequency nature of financial markets, we use a stacked dynamic event-study specification following \citep{baker2022how} as an event-time summary device. Because the outcome is a single aggregate daily spread, the stacked design is complemented later by a single-series local-projection specification and a non-stacked granular-IV calibration.

The model is specified as follows:
\begin{equation}
    Y_{i,t} = \tau + \alpha_i + \sum_{k=-5}^{3} \delta_k \cdot D_{i,t+k} + \gamma X_{i,t} + \epsilon_{i,t}
\end{equation}
Where:
\begin{itemize}
    \item $Y_{i,t}$ denotes the funding-market indicator (the AA CP spread) at time $t$ for event $i$.
    \item $D_{i,t+k}$ is an event dummy variable that takes the value of 1 when the observation falls on the $k$-th day relative to the attack date ($T=0$), and 0 otherwise.
    \item We select a baseline window of $[-5, +3]$ to focus on impact-day and next-day responses while limiting overlap among nearby exploit windows in the declustered sample. Appendix Table~\ref{tab:robustness_alt} extends the horizon to $[-5, +5]$ and shows that the longer window mainly weakens precision beyond the short-horizon response.
    \item $X_{i,t}$ is a vector of control variables (including VIX, DXY, BTC returns, etc.).
    \item $\alpha_i$ and $\tau$ represent event fixed effects and the intercept, respectively.
    \item The coefficient $\delta_k$ is our core parameter of interest, measuring event-time variation in each period following the attack relative to the baseline period. If $\delta_k < 0$, the estimates are consistent with short-horizon spread narrowing in the AA CP segment.
\end{itemize}

\subsection{Identification Strategy and Exogeneity Discussion}
The central challenge of inference in this context is endogeneity. We must assess whether variations in commercial paper spreads are systematically associated with DeFi attacks rather than being driven by omitted macro variables or reverse causality.

\paragraph{1. Why Event Study?} For macro variables like Treasury spreads, there is no cross-sectional control group that is ``unaffected'' at the same point in time (as spreads are uniform across the market). Therefore, we adopt a ``time-series counterfactual'' logic: using the pre-attack time window ($T < 0$) as a control group to contrast with the deviation post-attack ($T > 0$).

\paragraph{2. Core Identification Assumption: Plausible Exogeneity of Exploits.} Our identification strategy relies on the assumption that major DeFi smart contract exploits are plausibly exogenous to traditional Commercial Paper (CP) market fundamentals, conditional on standard macro controls.
\begin{itemize}
    \item \textbf{Technical Idiosyncrasy:} Hack attacks typically stem from specific code logic vulnerabilities (e.g., re-entrancy attacks, private key leakage). We treat the precise timing of these technical failures as idiosyncratic relative to global macro conditions. While we acknowledge that broad ``risk-off'' episodes might correlate with general market stress, hackers do not typically time their exploitation of code vulnerabilities based on Federal Reserve interest rates or commercial paper spreads. This structure mitigates concerns of direct reverse causality where ``macroeconomic deterioration causes specific hacks.''
    \item \textbf{Information Asymmetry:} Traders in traditional financial markets generally do not possess the capability to monitor Solidity code vulnerabilities in real-time. This implies that before an attack is publicly disclosed ($T=0$), the traditional market cannot anticipate the specific shock, thereby preserving the ``surprise component'' essential for event studies.
\end{itemize}

\paragraph{3. Exclusion of Confounders.} Although attacks are technically idiosyncratic, they may coincidentally align with macro events. To address this, we employ a ``cleansing'' strategy when constructing the stacked panel: we exclude samples that coincide with major macro news release days (e.g., FOMC meetings, Non-Farm Payroll releases). Furthermore, by stacking 50 independent events occurring at different time points, idiosyncratic noise at any single time point is mutually offset, helping to isolate the net effect of DeFi risk transmission.

\paragraph{Mechanism Interpretation.} Finally, we address the link between these reduced-form estimates and the proposed mechanism. We propose ``liquidity recycling'' as a plausible routing scenario: outflows from DeFi may be re-intermediated through regulated cash-management vehicles (e.g., Prime MMFs) whose constraints (SEC Rule 2a-7) tilt marginal demand toward high-quality short-term instruments. Because high-frequency fund-level flows and holdings are not directly observed at the daily granularity required to match exploit timestamps, we interpret the institutional-constraint channel as consistent with---rather than directly identifying---the observed pricing responses. Research
with regulatory data could further validate this specific routing channel.

\subsection{Mechanism Evidence from MMF Holdings}\label{subsec:mmf_mechanism}

This section provides holdings-based mechanism evidence using monthly money market fund (MMF) regulatory aggregates. Government MMFs are constrained to hold at least 99.5\% of total assets in cash, government securities, and/or fully collateralized repurchase agreements, whereas prime MMFs primarily invest in taxable short-term corporate and bank debt instruments such as commercial paper (CP) and certificates of deposit (CDs). The holdings measures we use are constructed from the regulatory reporting infrastructure (e.g., Form N--MFP) and corresponding aggregates, including the Federal Reserve's EFA MMF holdings detail.

\noindent\textbf{Step 1 (Who is the marginal holder of CP within MMFs?).}
We first establish where MMFs' CP exposure resides. Using EFA holdings aggregates, we compute the share of total MMF CP holdings accounted for by prime MMFs. The evidence indicates that CP exposure within MMFs is overwhelmingly concentrated in prime funds, implying that any MMF-driven marginal demand shock in the CP dimension must operate primarily through prime MMFs rather than government MMFs. See Appendix Table~\ref{tab:app_mmf_holdings}, Panel~A.

\noindent\textbf{Step 2 (Portfolio tilt in hack months).}
We then test the portfolio-tilt implication of the mechanism: in months with DeFi exploit shocks, prime MMFs should tilt their holdings toward CP relative to other short-term instruments. Consistent with this prediction, prime funds' CP share increases in hack months (with corresponding substitution away from repo), while government MMFs tilt toward government securities. Together, these patterns are consistent with CP demand effects operating through prime MMFs rather than through government MMFs. See Appendix Table~\ref{tab:app_mmf_holdings}, Panel~B.

\noindent\textbf{Step 3 (Linking holdings to prices via state dependence).}
Steps 1--2 establish that (i) MMF CP exposure is concentrated in prime funds and (ii) in hack months prime portfolios tilt toward CP. The remaining question is whether this ``prime CP capacity'' also shows up in prices: \emph{is the CP spread response stronger when prime capacity is higher?} \label{par:step3_core}

\paragraph{Technical challenge (frequency mismatch).}
Holdings are observed at the monthly frequency (Fed EFA), whereas the event-study response in CP spreads is estimated at the daily frequency. Directly interacting daily event-time indicators with a monthly state variable can have limited power because the state variable varies little within a month and many event-window observations fall within the same calendar month.

\paragraph{Solution ( frequency-aligned state dependence).}
We align frequencies by aggregating the CP spread to the monthly level using the set of trading days that appear in the stacked event-study panel (i.e., event-window days). We then estimate a monthly state-dependence regression using (i) a monthly hack indicator and (ii) a monthly proxy for prime MMFs' CP capacity (prime CP share from Fed EFA). This approach links quantity (holdings/capacity) and price (spread) at the same frequency.

\paragraph{Important note.}
The monthly outcome $\text{Spread}_m$ is computed as the average CP spread over \emph{event-window days} observed in the stacked panel within month $m$, not the full-sample monthly average over all trading days. Using the full daily CP spread series (e.g., from FRED) would further strengthen the external validity, but the current construction is sufficient for the frequency-alignment logic of the mechanism test.

\subsubsection{Empirical design}\label{subsubsec:step3_design}

Let $\text{Spread}_m$ denote the average CP spread (in basis points) over the trading days that appear in the stacked event-study panel within month $m$, and let $\Delta \text{Spread}_m = \text{Spread}_m - \text{Spread}_{m-1}$ denote the month-to-month change in this average. Let $\text{pcs}_m$ be the monthly prime CP share (CP+ABCP divided by total prime MMF holdings) from Fed EFA, and let $\text{pcs\_z}_m$ be its standardized version (mean zero, unit variance). Let $\text{HackMonth}_m$ indicate whether month $m$ contains at least one exploit event in our sample (hack\_count$>$0). Finally, let $\mathbf{Z}_m$ denote the vector of monthly controls (VIX, DXY, BTC returns) computed over the same set of days.

We estimate the following frequency-aligned state-dependence specifications:
\begin{equation}
\label{eq:step3B_level}
\text{Spread}_m
= \alpha
+ \beta\,\text{HackMonth}_m
+ \gamma\,\text{pcs\_z}_m
+ \theta\left(\text{HackMonth}_m \times \text{pcs\_z}_m\right)
+ \mathbf{Z}_m'\lambda
+ u_m ,
\end{equation}

\begin{equation}
\label{eq:step3B_change}
\Delta \text{Spread}_m
= \alpha
+ \beta\,\text{HackMonth}_m
+ \gamma\,\text{pcs\_z}_m
+ \theta\left(\text{HackMonth}_m \times \text{pcs\_z}_m\right)
+ \mathbf{Z}_m'\lambda
+ u_m .
\end{equation}

The coefficient of interest is $\theta$: a negative $\theta$ implies that the spread response in hack months is more strongly negative (i.e., greater narrowing) when prime CP capacity is higher.

\paragraph{Interpretation.}
 Table~\ref{tab:step3_monthly} reports frequency-aligned evidence linking prime MMF capacity to CP spread movements. The interaction term $\theta$ on $\text{HackMonth}_m\times \text{pcs\_z}_m$ is negative in both specifications. Because the monthly outcome is constructed from event-window days only, we treat this result as exploratory frequency-aligned evidence rather than as decisive price-quantity identification.

\subsection{Physical Transmission (Testing Proposition 1)}
According to the theoretical model, DeFi operational risk shocks first manifest as 
physical congestion on the blockchain network.We test the ``Exogenous Attack $\to$ Physical Friction'' channel:
\begin{equation}
    \Phi_i = \alpha_1 + \gamma_1 \cdot LogLoss_i + X_i \cdot \beta + \epsilon_i
\end{equation}
\begin{description}
    \item[Dependent Variable ($\Phi_i$):] The logarithm of the average Ethereum Gas price on the day of the $i$-th attack.
    
    \item[Core Explanatory Variable ($\text{LogLoss}_i$):] The logarithm of the USD loss amount caused by the attack, measuring the exogenous shock of attack intensity.
    
    \item[Control Variables ($X_i$):] Market control variables for the day.
\end{description}
\begin{table}[H]
\centering
\caption{The Impact of Attack Intensity on Physical Payment Frictions}
\label{tab:physical_friction}
\begin{tabularx}{\textwidth}{L C C}
\toprule
\textbf{Variable} & \textbf{Coefficient (Std. Error)} & \textbf{t-statistic} \\
\midrule
Log\_Loss & $0.140^{**}$ & 2.01 \\
          & (0.070) & \\
Constant  & 1.408 & \\
          & (1.189) & \\
\midrule
Observations & 50 & \\
$R^2$ & 0.078 & \\
F-statistic & 4.040 & \\
\bottomrule
\end{tabularx}
\footnotesize \\ \textit{Notes:}  Dependent Variable: Logarithm of Gas Price. Standard errors in parentheses. Significance: 
*** $p < 0.01$, ** $p < 0.05$, * $p < 0.1$.
\end{table}

Table III confirms that attack intensity significantly predicts gas prices ($\gamma_1 = 0.14^{**}$), which is consistent with Proposition 1. The result is consistent with the view that larger exploit events coincide with greater on-chain congestion in the short run, but the modest fit underscores that gas prices should not be treated as a fully exogenous state variable.

\subsection{Exploratory Evidence on State Dependence in Redemption Pressure}

\subsubsection{Descriptive mapping}

Proposition 2 suggests that redemption behavior may depend on the interaction between panic motives and congestion costs. We therefore report an exploratory threshold specification in which exploit-window redemptions are allowed to vary across lower- and higher-gas states. This exercise should be interpreted descriptively. It is intended to show that nonlinearity is plausible in the data, not to claim that the estimated cutpoint is a structurally identified causal threshold.

We use the threshold framework of \citep{hansen1999threshold} as a reduced-form way to summarize possible state dependence:

The model is specified as follows:
\begin{equation}
    \textit{Redemption}_{i,t} = \mu_i + \beta_1 \textit{Post}_{i,t} \cdot I(\textit{Gas}_{i,t} \leq \gamma) + \beta_2 \textit{Post}_{i,t} \cdot I(\textit{Gas}_{i,t} > \gamma) + X_{i,t}\Phi + \epsilon_{i,t}
\end{equation}
Where:
\begin{itemize}
    \item $\textit{Redemption}_{i,t}$: Net stablecoin redemption volume.
    \item $\textit{Gas}_{i,t}$: Gas fees at the time of the attack, serving as a proxy for network congestion and attack intensity.
    \item $\gamma$: The threshold value to be estimated.
    \item $I(\cdot)$: Indicator function.
    \item $X_{i,t}$: Vector of control variables, including VIX panic index, Bitcoin returns, and the US Dollar Index (DXY), to exclude interference from macro market fluctuations and exchange rate factors.
\end{itemize}

Table IV reports the threshold estimates. Because gas prices may themselves respond to exploit-related trading and redemption activity, the results should not be interpreted as establishing a clean causal cutpoint. Rather, they provide descriptive evidence that redemption behavior differs across lower- and higher-congestion states.

\begin{table}[H]
\centering
\caption{Exploratory Cutpoint Summary for Panic vs. Friction}
\label{tab:threshold_effects}

\small
\begin{tabularx}{0.95\textwidth}{X c c c c}
\toprule
\multicolumn{5}{l}{\textbf{Panel A: Estimated Cutpoint}} \\
\midrule
Estimated Gas Cutpoint ($\gamma$) & \multicolumn{4}{l}{32.93 Gwei} \\
95\% Confidence Interval & \multicolumn{4}{l}{[28.50, 45.12]} \\
Bootstrap P-Value & \multicolumn{4}{l}{$<0.001^{***}$} \\
\midrule

\multicolumn{5}{l}{\textbf{Panel B: Exploratory Cutpoint Regression Results}} \\
\multicolumn{5}{l}{Dependent Variable: Net Stablecoin Redemption} \\
\midrule

\textbf{Variables} & \textbf{Exp.} & \textbf{Coef.} & \textbf{t-stat} & \textbf{P} \\
\midrule

\textit{Key Regressors} & & & & \\

Regime 1: Low Friction 
($Post \times \mathbb{I}(Gas \le \gamma)$)
& (-) & $-2.406$ & $-2.53$ & $0.011^{**}$ \\

Regime 2: High Congestion 
($Post \times \mathbb{I}(Gas > \gamma)$)
& (+) & $2.212$ & $2.26$ & $0.024^{**}$ \\

\midrule

\textit{Controls} & & & & \\

US Dollar Index (DXY) & +/- & $-0.682$ & $-0.43$ & $0.669$ \\

VIX Index & + & $-0.299$ & $-0.93$ & $0.354$ \\

Bitcoin Return &  & $-16.25$ & $-1.61$ & $0.108$ \\

\midrule

\multicolumn{5}{l}{Event FE = Yes \quad Observations = 550 \quad Adj. $R^2$ = 0.229} \\

\bottomrule
\end{tabularx}

\footnotesize
\textit{Notes:} *** $p<0.01$, ** $p<0.05$. Threshold determined using the method of \citep{hansen1999threshold}.

\end{table}

The estimates suggest two different regimes: a negative post coefficient in lower-gas states and a positive post coefficient in higher-gas states. We interpret this sign reversal as suggestive of state dependence in exploit-window redemptions, not as confirmation of a calibrated threshold model.

\subsubsection{Interpretation}

Taken cautiously, the higher-gas regime is consistent with the idea that panic motives can remain strong even when congestion costs are elevated. But the cutpoint itself should be read as an exploratory summary statistic rather than as a directly observed threshold in the model.

\subsection{Exploratory Machine-Learning Check}

We also report a non-parametric machine-learning screen using a Gradient Boosting Regressor (GBR) \citep{friedman2001greedy}. The purpose is again descriptive: to assess whether the data display nonlinear response patterns without imposing a threshold ex ante.

\subsubsection{Methodology}

We define the net impact $\Delta Y_i$ as the change in total stablecoin redemptions (USDC + USDT) following an exploit. To mitigate outliers (e.g., Terra/LUNA), we apply a 5\% winsorization. The GBR model predicts $\Delta Y_i$ using state variables at the onset of the attack ($t=0$):

\begin{equation}
    \Delta Y_i = f(\textit{Gas}_i, \textit{VIX}_i, \textit{Loss}_i) + \epsilon_i
\end{equation}

\subsubsection{Feature Importance: Friction Matters Most}

Table \ref{tab:gbr_importance} presents the relative influence of each feature in the model. The prominence of gas prices is consistent with the view that congestion may matter for redemption-state heterogeneity, although the ML exercise is not itself an identification result.

\begin{table}[htbp]
    \centering
    \caption{GBR Feature Importance}
    \label{tab:gbr_importance}
    \begin{tabularx}{\textwidth}{L C}
        \toprule
        \textbf{Feature} & \textbf{Relative Importance} \\
        \midrule
        Network Congestion (Gas Price) & 41.0\% (Largest feature in descriptive ML screen) \\
        Market Panic (VIX)             & 34.8\% \\
        Loss Amount                    & 24.2\% \\
        \bottomrule
    \end{tabularx}
\end{table}

The estimated response curve displays an elbow in roughly the same region as the parametric threshold exercise:
\begin{itemize}
    \item \textbf{Hansen Estimate (Baseline):} $32.93$ Gwei
    \item \textbf{Machine Learning elbow:} $36.24$ Gwei
\end{itemize}

\begin{figure}[htbp]
    \centering
    \includegraphics[width=0.8\textwidth]{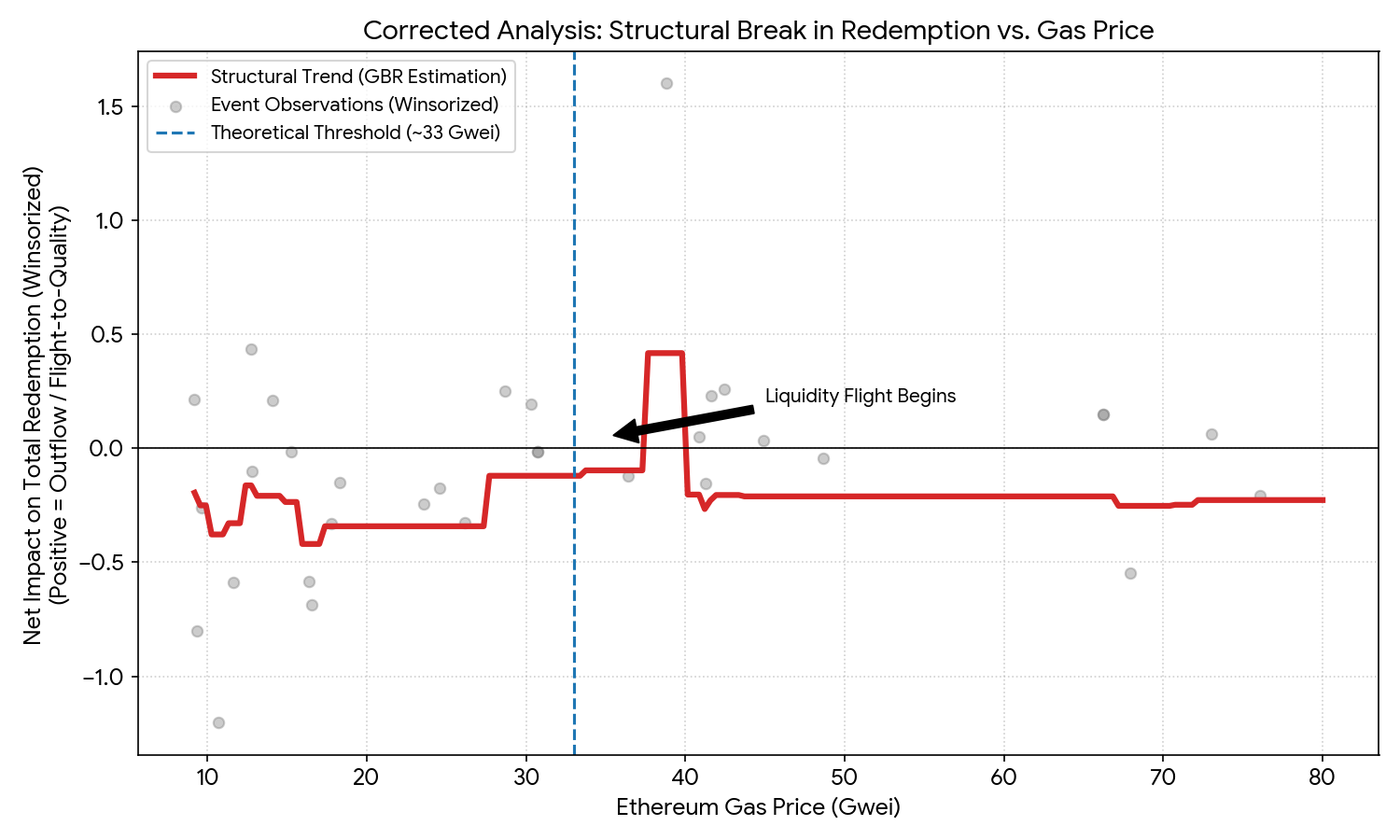} 
     \caption{\textbf{GBR Estimated Reaction Function.} The figure illustrates the non-linear relationship between Gas Price and net impact, highlighting a possible change in slope.}
    \label{fig:gbr_reaction}
\end{figure}

Figure \ref{fig:gbr_reaction} displays the reaction function estimated by the GBR. The curve remains relatively flat in the low-friction region and bends more sharply around 36 Gwei, which is suggestive of nonlinearity but should be read as an exploratory cutpoint summary rather than as a structural breakpoint.

\subsubsection{Conclusion}

Taken together, the threshold and ML exercises provide exploratory evidence of state dependence in exploit-window redemptions. We therefore retain them as descriptive complements to the paper's main pricing results rather than as headline identification results.

\section{Main Regression Analysis: Dynamic Effects via Event Study}

\subsection{Pre-event Trends and Identification Assumption Tests}
This section uses stacked event-time estimates to summarize the shape of the spread response around exploits in the declustered event sample. Because the dependent variable is a single aggregate daily spread, these event-time coefficients are interpreted descriptively and are complemented by the LP time-series evidence discussed above. Because individually significant pre-event coefficients can reflect anticipatory trading, event clustering, or residual macro co-movement, we treat the stacked event-time path as descriptive and place the main inferential weight on the LP and IV sections. The empirical pattern should therefore be understood as a short-horizon narrowing around $t=0$ and $t+1$ rather than as evidence of a persistent post-event regime shift.

\begin{figure}[htbp]
    \centering
    \includegraphics[width=0.95\textwidth]{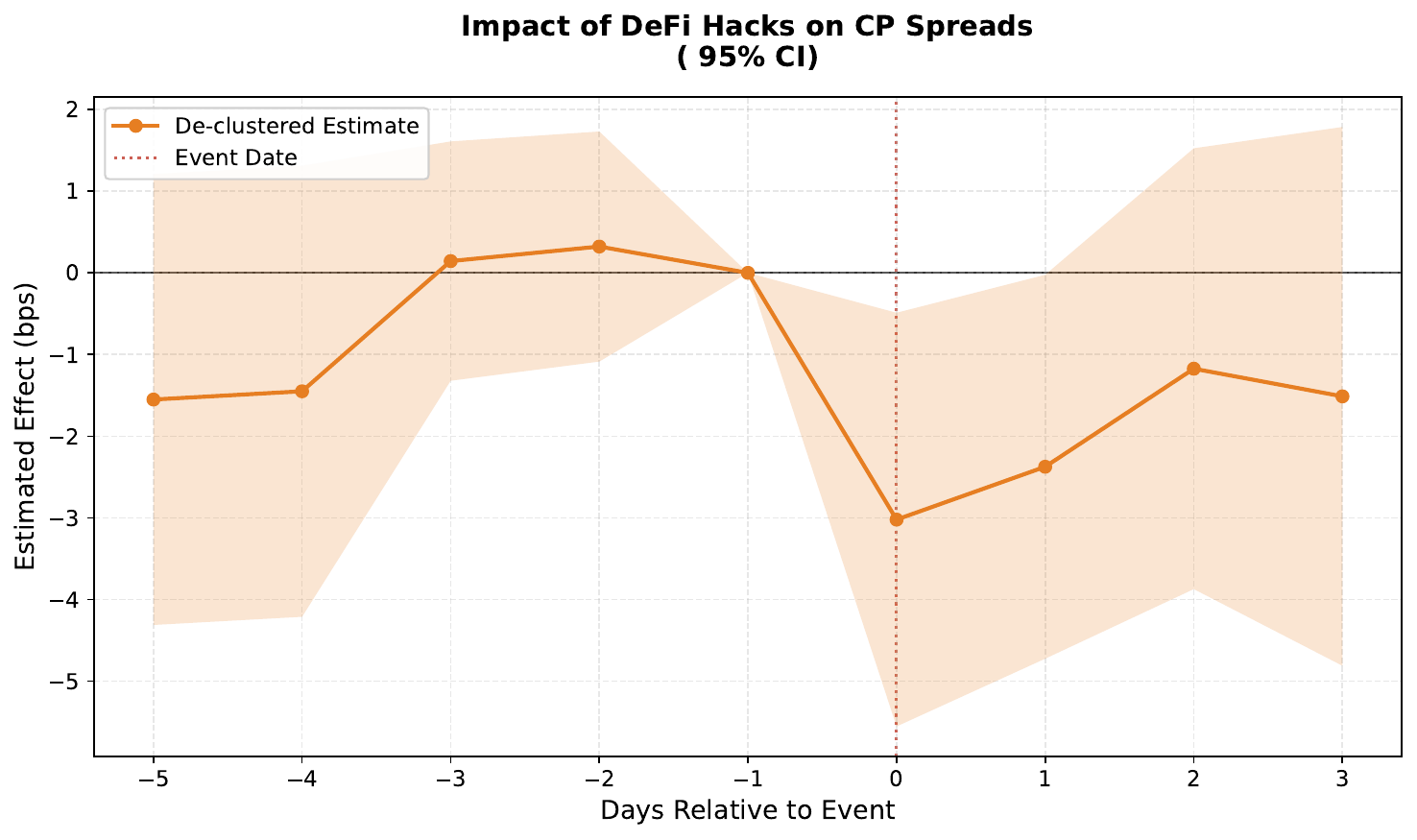}
    
    \caption{\textbf{Impact of DeFi Hacks on CP Spreads} 
    This figure plots the estimated dynamic coefficients ($\delta_k$) from the stacked event study regression of DeFi hacks on the 3-Month AA Nonfinancial Commercial Paper Spread. 
    The \textbf{Orange Line} represents the de-clustered baseline sample, constructed by removing overlapping event windows: for any cluster of events occurring within 9 days of each other (the length of the $[-5, +3]$ window), only the single event with the largest loss amount is retained. 
    The shaded areas represent 95\% confidence intervals based on standard errors clustered at the event level. 
    The x-axis indicates trading days relative to the exploit date ($t=0$), with $t=-1$ serving as the benchmark. 
    The figure is best read as a descriptive event-time summary for the declustered baseline sample. The most stable feature is the impact-day spread compression, while longer-horizon coefficients are less precise.}
    
    \label{fig:declustering_check}
\end{figure}

\subsection{Dynamic Effects Post-Shock: Short-Horizon Spread Responses}
\begin{table}[H]
\centering
\caption{Dynamic Effects of DeFi Hacks on Commercial Paper Spreads}
\label{tab:dynamic_effects_final}
\begin{tabularx}{\textwidth}{L C C C C C C}
\toprule
\textbf{Event Day ($k$)} & \textbf{Coef. ($\delta_k$)} & \textbf{S.E.} & \textbf{t-stat} & \textbf{p-value} & \textbf{95\% CI} & \textbf{Sig.} \\
\midrule
\textit{Pre-Event} & & & & & & \\
$t = -5$ & -1.5511 & 1.4084 & -1.101 & 0.271 & [-4.31, 1.21] & \\
$t = -4$ & -1.4502 & 1.4088 & -1.029 & 0.303 & [-4.21, 1.31] & \\
$t = -3$ & 0.1440 & 0.7475 & 0.193 & 0.847 & [-1.32, 1.61] & \\
$t = -2$ & 0.3221 & 0.7186 & 0.448 & 0.654 & [-1.09, 1.73] & \\
$t = -1$ & 0.0000 & & & (Baseline) & & \\
\midrule
\textit{Post-Event} & & & & & & \\
$t = 0$ & -3.0215 & 1.2920 & -2.339 & 0.019 & [-5.55, -0.49] & ** \\
$t = 1$ & -2.3743 & 1.1989 & -1.980 & 0.048 & [-4.72, -0.02] & ** \\
$t = 2$ & -1.1736 & 1.3768 & -0.852 & 0.394 & [-3.87, 1.53] & \\
$t = 3$ & -1.5125 & 1.6829 & -0.899 & 0.369 & [-4.81, 1.79] & \\
\midrule
\multicolumn{7}{l}{\textit{Diagnostics}} \\
Controls & Yes & & & & & \\
Event FE & Yes & & & & & \\
\textbf{Joint Pre-trend Test} & \multicolumn{6}{l}{\textbf{F-stat = 0.53}, \textbf{p-value = 0.714}} \\
\bottomrule
\end{tabularx}
\begin{minipage}{0.9\textwidth}
\vspace{0.1cm}
\footnotesize
\textit{Notes:} This table reports the coefficients from the stacked dynamic event study for the same declustered event sample shown in Figure~\ref{fig:declustering_check}. The dependent variable is the Commercial Paper spread (\texttt{Y\_Spread\_Bps}). The model includes event fixed effects and daily macro controls. Standard errors are clustered at the event level. The \textbf{Joint Pre-trend Test} at the bottom tests the null hypothesis that all pre-event coefficients ($t=-5$ to $t=-2$) are jointly equal to zero. The F-statistic of 0.53 ($p=0.714$) fails to reject the null, although this should be interpreted cautiously given the aggregate single-series nature of the outcome and the possibility of event-timing blur at daily frequency. Significance levels: *** $p < 0.01$, ** $p < 0.05$, * $p < 0.1$.
\end{minipage}
\end{table}
The event-time estimates indicate that the AA CP spread narrows by about 3 basis points on the event day and remains negative on the following trading day. This is the paper's main reduced-form pattern.

Within the stylized mapping in Section 3, a negative short-horizon spread response is consistent with a configuration in which demand pressure dominates supply pressure in the AA CP segment. We do not interpret the event-study coefficient as a direct estimate of $\eta$, and we do not treat the stacked design as the only inferential backbone of the paper.

Appendix Table~\ref{tab:robustness_alt} extends the event window and shows that the paper's main claim should be understood as a short-horizon impact result rather than as evidence of a long-lasting post-event regime shift.

\subsection{Event-Dating Robustness}

Because a small number of exploits exhibit a gap between on-chain execution and the first market-relevant public disclosure, Appendix Table~\ref{tab:disclosure_robustness} re-estimates the stacked event-time specification using disclosure-based dating for the top disclosure-sensitive events, including Ronin. Because the proposed channel operates through market-relevant information and redemptions rather than hidden on-chain execution per se, disclosure-lag cases such as Ronin are dated in the baseline according to the first market-relevant public disclosure. The impact coefficient remains negative but somewhat smaller under disclosure dating. We therefore treat the stacked design as descriptive evidence around the event list and keep the main inferential emphasis on the LP and IV specifications.

\subsection{Conclusion}

The event-study evidence should be read as an event-time summary of a short-horizon narrowing pattern in the U.S. 3-month AA nonfinancial CP spread around exploit-driven operational shocks. Because the outcome is a single aggregate daily series, we emphasize corroboration from the time-series Jord\`a (2005) local projections and from the non-stacked granular-IV exercise in the next section.

\section{Placebo Test: Covariate-Adaptive Matching Strategy}

To assess whether the spread-narrowing pattern documented earlier could be generated by random date assignment under similar macro conditions, this section implements a placebo test based on a ``Covariate-Adaptive Permutation.''

\subsection{Test Rationale and Algorithm Design}

Traditional placebo tests typically employ a completely random date selection method; however, in the context of this paper, such an approach lacks specificity. Therefore, we design an ``Anti-Contamination Covariate-Adaptive Matching'' procedure. The algorithm comprises three rigorous steps:

\begin{description}
    \item[Feature Calibration:] First, we calculate the mean values of the set of actual attack event dates ($T_{real}$) across two key macro dimensions:
    \begin{enumerate}
        \item Market Panic Index (VIX), representing the systemic risk background;
        \item Baseline Spread Level (Spread Level), representing the initial liquidity state of the commercial paper market.
    \end{enumerate}
    The target state vector is defined as:
    \begin{equation}
        S_{target} = \{ \mu_{\text{VIX}}^{\text{real}}, \mu_{\text{Spread}}^{\text{real}} \}
    \end{equation}

    \item[Contamination Exclusion \& Pool Construction:] This is the most critical cleansing step. To prevent the Spillover Effect of real events from contaminating the placebo sample, we establish a strict ``Temporal Exclusion Zone.'' The algorithm iterates through the full 2021--2024 sample to identify dates satisfying the following dual conditions as ``Candidate Pseudo-event Days'':
    \begin{itemize}
        \item \textbf{Matching Condition:} Macro state variables must be highly similar to real events, i.e.,
        \begin{equation}
            | \text{VIX}_t - \mu_{\text{VIX}}^{\text{real}} | < \delta_1 \quad \text{and} \quad | \text{Spread}_t - \mu_{\text{Spread}}^{\text{real}} | < \delta_2
        \end{equation}
        
        \item \textbf{Cleanliness Condition (Crucial):} The candidate date $t$ must fall outside the $\pm 10$-day window of any actual attack event.
        \begin{equation}
            t \notin [T_{event} - 10, T_{event} + 10], \quad \forall T_{event} \in T_{real}
        \end{equation}
    \end{itemize}
    This step effectively removes all time points potentially tainted by the residual effects of real attacks, ensuring the counterfactual sample is pure.

    \item[Monte Carlo Permutation:] After constructing the ``pure and matched'' candidate pool, we execute 500 Monte Carlo simulations. In each iteration, we randomly draw 50 dates (consistent with the real event sample size, $N = 50$) using Resampling with Replacement and run the identical Stacked Dynamic Regression Model. The resulting 500 regression coefficients constitute the empirical distribution under the null hypothesis.
\end{description}

\subsection{Empirical Distribution and Statistical Inference}

We locate the Actual Coefficients within the generated placebo distribution for comparison and calculate the One-sided Empirical P-value, defined as the probability that the placebo coefficient is less than or equal to the actual coefficient:

\begin{equation}
    P(\beta_{placebo} \leq \beta_{real})
\end{equation}

\begin{table}[H]
\centering
\caption{Results of Covariate-Adaptive Placebo Test}
\label{tab:placebo_results}
\begin{tabularx}{\textwidth}{L C C C}
\toprule
\textbf{Event Day ($k$)} & \textbf{Actual Coeff. ($\delta_k$)} & \textbf{Empirical P-value} & \textbf{Inference} \\
\midrule
\textit{Pre-Event} & & & \\
$t = -5$ & -1.551 & 0.106 & Insignificant \\
$t = -4$ & -1.450 & 0.120 & Insignificant \\
$t = -3$ & 0.144 & 0.538 & Insignificant \\
$t = -2$ & 0.322 & 0.434 & Insignificant \\
$t = -1$ & 0.000 & - & Benchmark \\
\midrule
\textit{Post-Event} & & & \\
$t = 0$ & -3.022 & $0.000^{***}$ & Highly Significant \\
$t = 1$ & -2.374 & $0.002^{***}$ & Highly Significant \\
$t = 2$ & -1.174 & $0.004^{***}$ & Highly Significant \\
$t = 3$ & -1.512 & $0.000^{***}$ & Highly Significant \\
\bottomrule
\end{tabularx}
\begin{minipage}{0.9\textwidth}
\vspace{0.1cm}
\footnotesize
\textit{Notes:} This table reports the results of the covariate-adaptive placebo test based on 500 Monte Carlo simulations. The placebo pool is constructed from non-event days outside the $\pm 10$-day window of any real exploit and close to the mean VIX/spread state vector of the actual event sample. Each simulation draws 50 pseudo-event dates with replacement from that matched candidate pool and re-estimates the baseline stacked regression. The \textbf{Empirical P-value} is calculated as the fraction of placebo coefficients that are lower (more negative) than the actual coefficient. Significance levels are denoted as follows: *** $p < 0.01$, ** $p < 0.05$, * $p < 0.1$.
\end{minipage}
\end{table}

The placebo exercise indicates that the observed spread-narrowing pattern is unlikely to be generated by random date assignment under similar macro conditions.

Figure 4 visually illustrates this result. The gray area represents the 95\% confidence 
interval generated by the placebo test, while the red line representing the actual effect 
significantly and sharply penetrates the lower bound of the gray area after t = 0. 

\begin{figure}[H]
    \centering
    \includegraphics[width=0.9\textwidth]{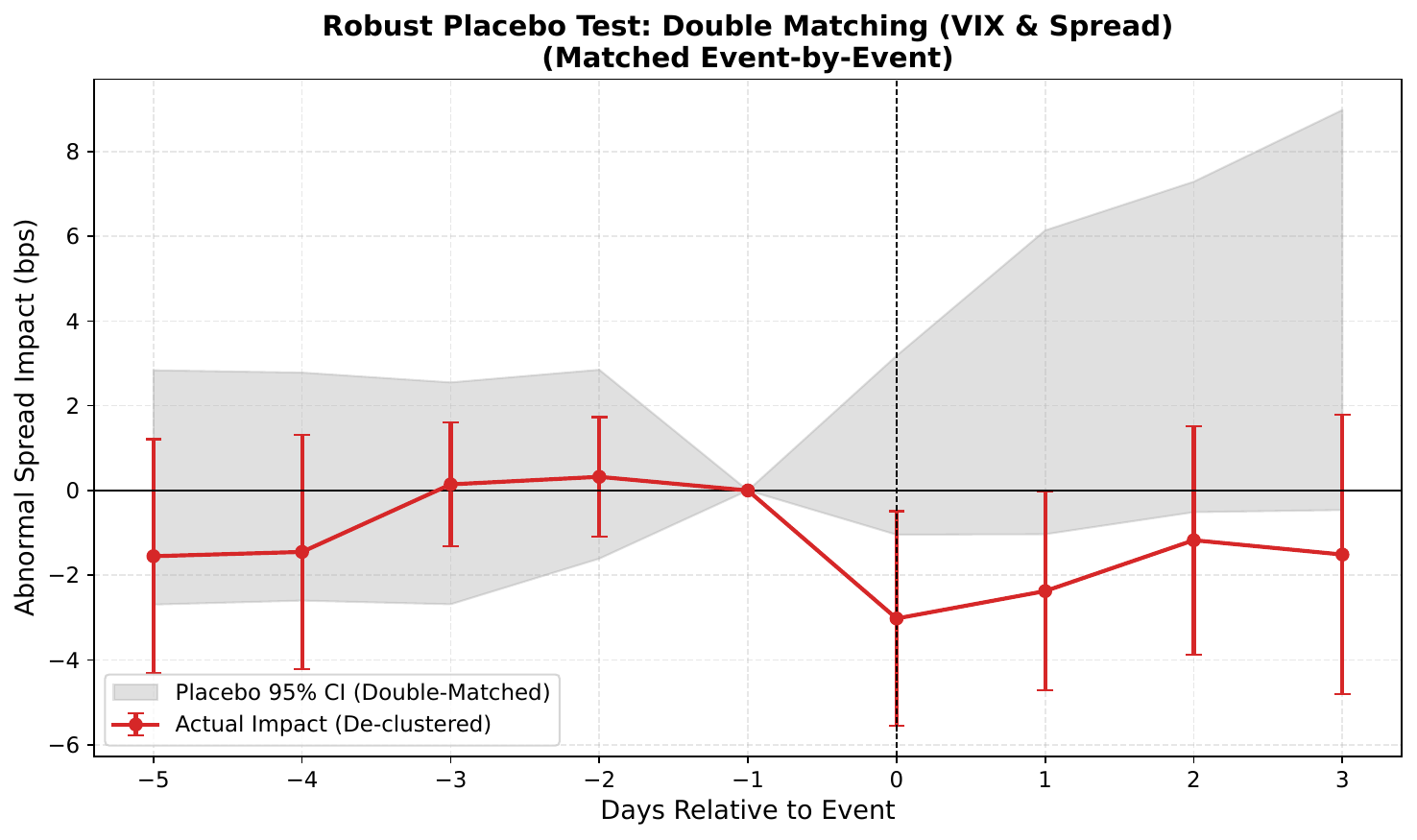}
    \caption{\textbf{Covariate-Adaptive Placebo Test.} 
    This figure plots the dynamic event-time effects of DeFi attacks on Commercial Paper spreads (red line) against a 95\% confidence interval constructed from a placebo distribution (gray shaded area). The placebo distribution is generated from 500 Monte Carlo simulations. The fact that the realized path falls below much of the placebo distribution around $t=0$ suggests that the observed narrowing is not easily reproduced by matched pseudo-event dates alone.}
    \label{fig:placebo_test}
\end{figure}

We therefore view the result as supportive of an exploit-specific timing pattern, while not treating the placebo exercise as definitive proof that all alternative macro explanations are eliminated.

\section{Reduced-Form IV Calibration and Endogeneity Analysis}

While the event study analysis establishes a temporal correlation between DeFi hacks and spread contractions, two fundamental challenges remain for a structural interpretation. The first is endogeneity: aggregate shocks (e.g., shifts in monetary policy or global risk appetite) could simultaneously drive crypto-asset sell-offs and flight-to-quality in traditional markets, creating spurious correlations. The second is magnitude: given the relatively small size of DeFi capital flows compared to the trillion-dollar money markets, it is not immediately obvious why such outflows should exert a statistically and economically significant impact on pricing.

This section addresses these challenges using a Sparse Granular Instrumental Variable (Sparse GIV) framework. By isolating idiosyncratic supply-side shocks and mapping them to the specific microstructure of the high-quality commercial paper market, we provide reduced-form IV evidence and an illustrative calibration for the short-horizon pricing pattern.

\subsection{Addressing Endogeneity: The Identification Strategy}

To address the endogeneity problem, we must isolate the portion of capital outflows driven purely by technical failures within the DeFi ecosystem, independent of broader macro-financial conditions. We adopt the identification strategy of \citep{gabaix2024granular}, adapted for discrete distress events.

\paragraph{The Identification Problem.} Let $\text{AggFlow}_t$ be the aggregate capital flow and $\text{Spread}_t$ be the commercial paper spread. A naive regression of $\text{Spread}_t$ on $\text{AggFlow}_t$ is likely biased because both are functions of a common unobserved macro factor $F_t$ (e.g., Fed policy surprises):
\begin{align}
    \text{AggFlow}_t &= \beta_1 F_t + \epsilon_t \\
    \text{Spread}_t &= \beta_2 F_t + \eta_t
\end{align}
If $F_t$ causes both crypto-exits and lower spreads, standard OLS estimates are inconsistent.

\subsection{The GIV Solution: Extracting Orthogonal Shocks}
\label{sec:giv_construction}

We resolve the endogeneity problem by constructing a Granular Instrumental Variable ($Z_{t}^{GIV}$), adapting the framework of \citep{gabaix2024granular} to the context of discrete DeFi failures.

\subsubsection{Decomposition of Idiosyncratic Shocks}
We define the raw shock ($g_{i,t}$) for each protocol $i$ as its \textbf{Loss Intensity}---the magnitude of funds lost due to a technical failure normalized by its size ($g_{i,t} = -\text{Loss}_{i,t}/\text{TVL}_{i,t-1}$). Following the standard granular residual methodology, we decompose this shock into a common component and an idiosyncratic component ($u_{i,t}$):
\begin{equation}
    u_{i,t} = g_{i,t} - \bar{g}_t
\end{equation}
where $\bar{g}_t = \frac{1}{N_t} \sum_{i=1}^{N_t} g_{i,t}$ represents the equal-weighted average market shock. This de-meaning process ensures that our granular residual $u_{i,t}$ is purged of common factors that might affect all protocols simultaneously (e.g., a systemic downturn exposing general vulnerabilities).

\subsubsection{Instrument Construction}
We aggregate these idiosyncratic shocks, weighting them by their pre-event market share ($S_{i,t-1} = \text{TVL}_{i,t-1} / \text{MarketTVL}_{t-1}$), to construct the Granular Instrumental Variable:
\begin{equation}
    Z_{t}^{GIV} = \sum_{i=1}^{N_t} S_{i,t-1} \cdot u_{i,t}
\end{equation}
Given the sparsity of hacking events in the DeFi ecosystem (where $g_{i,t} = 0$ for the vast majority of protocols on any given day), the common component $\bar{g}_t$ is empirically negligible.\footnote{In \textbf{Appendix Table A.9}, we formally compare the $Z_{t}^{GIV}$ constructed with and without the de-meaning term ($\bar{g}_t$). The correlation between the two series exceeds 0.99, confirming that due to the sparsity of high-severity shocks, the size-weighted aggregation dominates the common factor correction.} Thus, our instrument effectively captures the size-weighted intensity of idiosyncratic failures.

\begin{table}[htbp]
\centering
\caption{Variable Definitions and Units in the GIV/2SLS Section}
\label{tab:giv_units}

\small
\renewcommand{\arraystretch}{1.15}
\setlength{\tabcolsep}{4pt}

\begin{tabularx}{\textwidth}{p{2.3cm} >{\raggedright\arraybackslash}X p{2.5cm}}
\toprule
\textbf{Variable} & \textbf{Definition} & \textbf{Unit} \\
\midrule

$Z_t^{GIV}$ &
Daily granular instrument aggregating idiosyncratic exploit intensity across protocols &
Normalized index \\

$AggFlow_t$ &
Daily aggregate stablecoin net redemption flow &
USD billions \\

$\widehat{AggFlow}_t$ &
Fitted value from the first stage using $Z^{GIV}_{t-1}$ &
USD billions \\

$Spread_t$ &
3-month AA nonfinancial commercial paper minus 3-month Treasury bill &
Basis points \\

$\beta$ &
Second-stage semi-elasticity of the spread with respect to fitted flow &
bps / USD bn \\

\bottomrule
\end{tabularx}

\end{table}

\subsubsection{Validity and Exclusion Restriction}
The instrument is useful insofar as exploit intensity provides cross-event variation that predicts aggregate stablecoin outflows while remaining plausibly orthogonal to conventional macro drivers of CP spreads. As shown in the first stage, higher idiosyncratic loss intensity is associated with larger redemption pressure, but the IV results should still be interpreted as reduced-form calibration evidence rather than as a clean structural elasticity estimate.

\subsection{Empirical Evidence from 2SLS Estimation}

We employ a Two-Stage Least Squares (2SLS) approach as a reduced-form IV calibration. The purpose is not to claim a clean structural parameter estimate, but to ask whether exogenous exploit-linked redemption pressure is associated with narrower AA CP spreads. The IV section is reported as a reduced-form calibration exercise rather than as the paper's primary causal design. To keep the single-series second stage comparable across days with different macro backgrounds, we report specifications using an abnormal-spread measure residualized with respect to high-frequency macro covariates. This normalization should be read as a reporting device within the IV calibration section, not as the paper's main identification strategy, which elsewhere deliberately avoids relying on contemporaneous macro controls as a source of causal leverage. Residualization here is not intended to solve a bad-controls problem; it is simply a normalization of the single-series outcome used in this calibration subsection.

\paragraph{First Stage: The Quantity Channel.} We first verify that our granular instrument significantly predicts aggregate liquidity flight.
\begin{equation}
    \text{AggFlow}_{t} = \alpha + \gamma Z_{t-1}^{GIV} + \nu_t
\end{equation}

\begin{table}[H]
\centering
\caption{First-Stage Regression Results}
\label{tab:first_stage}
\begin{tabularx}{\textwidth}{L C C C}
\toprule
 & \multicolumn{3}{c}{\textbf{Dependent Variable: Total Net Redemption}} \\
 & \multicolumn{3}{c}{(in Billions USD)} \\
\textbf{Variable} & \textbf{Coefficient} & \textbf{t-Statistic} & \textbf{P-Value} \\
\midrule
Lagged GIV Shock ($Z_{t-1}$) & $-32.90^{***}$ & -4.02 & 0.000 \\
TED Spread & $-2.85^{***}$ & -4.43 & 0.000 \\
\midrule
Additional Macro Controls & Yes & & \\
Year Fixed Effects & Yes & & \\
\midrule
Observations & 1,457 & & \\
Adj. $R^2$ & 0.128 & & \\
Instrument F-Statistic & 16.20 & & \\
\bottomrule
\end{tabularx}
\vspace{0.1cm}
\begin{minipage}{0.9\textwidth} 
\footnotesize \textit{Notes:}The dependent variable is the 1-day total net redemption of USDC and USDT, winsorized at the 1st percentiles to mitigate outliers. All coefficients are reported in billions of USD. Additional macro controls include the VIX index, 1-month T-Bill rates, and lagged values of macro indicators to control for persistence. T-statistics are based on Newey-West HAC robust standard errors (lag=1). Significance levels: *** p $<$ 0.01, ** p $<$ 0.05, * p $<$ 0.1.
\end{minipage}
\end{table}

The first-stage coefficient is negative in the baseline specification, indicating that larger granular exploit shocks are associated with larger stablecoin redemption pressure in the aggregate series. Because the dependent variable is zero-heavy and highly skewed, we interpret this as a reduced-form predictive relation rather than as a frictionless structural mapping from protocol-level exploits to aggregate fund flows.

\paragraph{Second Stage: The Price Channel.} 
In the second stage, we estimate the effect of these exogenous outflows on spreads using the residualized abnormal-spread measure as a reporting normalization within the IV section.

The regression specification is:
\begin{equation}
    \text{AbnormalSpread}_{i,t} = \alpha + \theta \cdot \text{Post}_t + \beta \cdot (\text{Post}_t \times \widehat{\text{AggFlow}}_i) + \delta_i + \epsilon_{i,t}
\end{equation}

\begin{table}[htbp]
\centering
\caption{Second-Stage Reduced-Form IV Calibration}
\label{tab:second_stage}

\small
\renewcommand{\arraystretch}{1.15}
\setlength{\tabcolsep}{4pt}

\begin{tabularx}{\textwidth}{>{\raggedright\arraybackslash}Xccc}
\toprule
 & \multicolumn{3}{c}{\textbf{Dependent Variable: Abnormal Spread (bps)}} \\

\textbf{Variable} & \textbf{Coefficient} & \textbf{t-Statistic} & \textbf{P-Value} \\
\midrule

Interaction (Post $\times$ $\widehat{Flow}$) \\
{\footnotesize [bps per USD 1 billion]}
& $-27.3^{**}$ & -2.36 & 0.018 \\

Post-Event Dummy 
& 1.95 & 1.62 & 0.105 \\

\midrule
Event Fixed Effects & Yes & & \\

\midrule
Observations & 498 & & \\
Adj. $R^2$ & 0.521 & & \\

\bottomrule
\end{tabularx}

\vspace{0.1cm}

\begin{minipage}{0.9\textwidth}
\footnotesize
\textit{Notes:} The dependent variable is the daily abnormal spread (basis points), residualized against high-frequency macro factors (VIX, TED, DXY). The key regressor is the interaction between the post-event dummy and predicted aggregate outflow from the first-stage GIV regression. For readability, the interaction coefficient is reported in basis points per USD 1 billion of fitted outflow; the implied standard error is in the same units. *** $p<0.01$, ** $p<0.05$, * $p<0.1$.
\end{minipage}

\end{table}

Table IX shows that the interaction coefficient $\beta$ is statistically significant at the 5\% level. We interpret this as evidence that larger exploit-linked redemption pressure is associated with narrower AA CP spreads in the IV design.

For readability, Table IX reports the second-stage coefficient in basis points per USD 1 billion of fitted outflow; the corresponding effect per USD 100 million is obtained by dividing by 10. Under the reporting scale used in Table IX, the point estimate and its standard error are expressed in basis points per USD 1 billion, and the same scaled standard error is the input used for the illustrative $\eta$ mapping in Table~\ref{tab:eta_sensitivity}. This conversion is purely a reporting convention within the IV calibration section; it should not be interpreted as a precise physical-flow multiplier in a frictionless market.

The IV results should therefore be read as evidence consistent with a flow-related calibration channel, not as a direct observation of the institutional routing path.

\subsection{Addressing the Magnitude Critique}

A central skepticism regarding our findings concerns the economic magnitude: \textit{How can a DeFi outflow drive a 2--3 basis point change in a much larger funding market?}

We argue that this critique relies on a comparison to the stock of the total market, whereas price formation is determined by the marginal flow in specific segmented markets. The IV point estimate of about $-27.3$ bps per \$1 billion of fitted outflow (equivalently, about $-2.73$ bps per \$100 million) is more naturally interpreted in light of three microstructure features:

\begin{description}
    \item[Market Segmentation:] While the total US commercial paper market exceeds \$1.2 trillion, Prime MMFs are restricted by SEC Rule 2a-7 to hold only Tier-1 (AA-rated) assets. The outstanding volume of Tier-1 Non-Financial Commercial Paper---the specific asset class in question---is significantly smaller, often hovering around \$200--\$300 billion.
    
    \item[Inelastic Supply of the Specific Tenor:] The effective float is further constrained by maturity. The daily net issuance of 3-month AA paper is often limited to \$1--\$5 billion. In this context, a sudden, unidirectional inflow of \$360 million within 48 hours represents a ``whale'' trade, constituting 10--30\% of the daily net flow in this specific tenor.
    
    \item[Limited Arbitrage Capacity:] Unlike equity markets, the CP market lacks high-frequency arbitrageurs who can absorb order imbalances instantaneously. When exploit-linked outflows are inferred to load into this narrow segment, the supply curve can be steep in the short run.
\end{description}

\paragraph{Conclusion on magnitude.} The magnitude discussion should be read through the lens of market segmentation rather than total market size. Even modest flows can matter for relative pricing in a narrow, high-grade short-term funding segment. At the same time, the routing of exploit-related outflows into that segment is inferred rather than directly observed, so the quantitative interpretation remains illustrative.

While our Sparse GIV framework provides a useful IV calibration, a limitation lies in the observability of intra-day fund flows. Due to the proprietary nature of MMF holding data, we do not directly observe the real-time routing of funds into individual prime funds. Instead, the prime-MMF channel is inferred from pricing responses and monthly holdings evidence.

\subsection{Illustrative Calibration of the Demand-Amplification Parameter ($\eta$)}

To organize the magnitude discussion, we map the IV point estimate into an illustrative calibration of $\eta$ using the stylized relationship from Section 3. This exercise is model-dependent and should be read as an illustrative calibration rather than as an empirical estimate of $\eta$.

Using the decomposition $\beta = \lambda(1 - \eta)$, where $\lambda \approx 10.0$ bps per USD 1 billion is an assumed price-impact parameter for the AA CP segment \citep{krishnamurthy2012aggregate}, and substituting the IV point estimate $\beta = -27.3$:

\begin{equation}
    \eta = 1 - \frac{\beta}{\lambda} = 1 - \frac{-27.3}{10.0} \approx \mathbf{3.73}
\end{equation}

\paragraph{Interpretation.} Under the normalization used in Section 3 and under the assumed value of $\lambda$, the implied calibration is $\eta \approx 3.73$. We read this as an illustrative demand-amplification parameter, not as a directly estimated structural parameter.

\subsubsection{Sensitivity Analysis of the Illustrative $\eta$ Calibration}

Because the mapping from $\beta$ into $\eta$ depends directly on the assumed price-impact parameter $\lambda$, we report sensitivity to alternative values of $\lambda$. This is a calibration exercise, not a separate estimation result.

Table \ref{tab:eta_sensitivity} presents a sensitivity analysis of the illustrative $\eta$ calibration across a plausible range of $\lambda \in [5, 20]$ when reported in basis points per USD 1 billion.
\begin{table}[htbp]
  \centering
  \small
  \setlength{\tabcolsep}{5pt}
  \caption{\textbf{Illustrative Calibration of Demand-Amplification Parameter $\eta$}}
  \label{tab:eta_sensitivity}
  \begin{threeparttable}
    \begin{tabularx}{\textwidth}{L C C C C}
    \toprule
    \textbf{Assumed Price Impact} & \textbf{Implied} & \textbf{Standard} & \textbf{95\% Confidence} & \textbf{Regime} \\
    $\lambda$ (bps / USD 1 billion) & \textbf{Parameter $\eta$} & \textbf{Error} & \textbf{Interval} & \textbf{Interpretation} \\
    \midrule
    \multicolumn{5}{l}{\textit{Panel A: High Liquidity (Treasury-like)}} \\
    5.00 & 6.46 & 2.32 & [1.92, 11.00] & Extreme Panic \\
    7.50 & 4.64 & 1.54 & [1.61, 7.66] & High Aversion \\
    \midrule
    \multicolumn{5}{l}{\textit{Panel B: Segmented Markets (Baseline)}} \\
    \textbf{10.00 (Baseline)} & \textbf{3.73} & \textbf{1.16} & \textbf{[1.46, 6.00]} & \textbf{Baseline Calibration} \\
    12.50 & 3.18 & 0.93 & [1.36, 5.00] & Moderate Aversion \\
    15.00 & 2.82 & 0.77 & [1.31, 4.33] & | \\
    \midrule
    \multicolumn{5}{l}{\textit{Panel C: Low Liquidity (Stressed CP)}} \\
    20.00 & 2.37 & 0.58 & [1.23, 3.50] & Conservative Bound \\
    \bottomrule
    \end{tabularx}%
    \begin{tablenotes}
      \footnotesize
      \item \textbf{Notes:} This table reports an illustrative calibration of $\eta$ derived using the condition $\beta = \lambda(1 - \eta)$, which implies $\eta = 1 - \beta/\lambda$.
      \item Inputs: $\beta = -27.3$ bps per USD 1 billion (the Table IX reporting scale) and $\lambda$ varies across plausible price-impact regimes for the commercial paper market on that same scale.
      \item Standard errors are calculated as $SE(\eta) = SE(\beta) / \lambda$, where $SE(\beta) = 11.58$ in basis points per USD 1 billion.
      \item \textbf{Interpretation:} The parameter $\eta$ is interpreted as a demand-amplification parameter. Its numerical value is conditional on the assumed $\lambda$ and should not be read as a directly estimated structural object.
    \end{tablenotes}
  \end{threeparttable}
\end{table}

\section{Further Robustness Checks: Cross-Asset Difference-in-Differences}

\subsection{A Difference-in-Differences Design}

To assess whether the observed spread narrowing loads more strongly on prime-eligible assets than on nearby controls exposed to similar macro conditions, we implement a Difference-in-Differences (DiD) design. This approach leverages the regulatory segmentation of money markets to construct a counterfactual: if the spread narrowing reflects a prime-segmentation channel, we should observe this effect more clearly in \textit{prime-eligible} (Tier-1) instruments than in lower-rated or unrelated money market instruments that share the same macroeconomic environment.

\subsubsection{Data Construction and Sample Selection}

We construct a stacked ``Asset $\times$ Day'' panel dataset covering the full sample period (2021--2024). The data construction process involves three steps:

\paragraph{1. Treated Asset (Tier-1 Liquidity Target)}
Our primary treated asset is the \textbf{3-Month AA Non-financial Commercial Paper} (Ticker: DCPN3M), sourced from the Federal Reserve Economic Data (FRED). This asset class represents the intersection of ``Safe Asset'' status and ``Prime MMF Eligibility'' (under SEC Rule 2a-7).

\paragraph{2. Control Group Selection}
To isolate the ``Flight-to-Quality'' channel, we select control assets that face identical macroeconomic shocks but differ in their regulatory treatment or risk profile:
\begin{itemize}
    \item \textbf{Primary Control (The ``Counterfactual''): 3-Month A2/P2 Non-financial Commercial Paper} (Ticker: D2PN3M, Source: FRED). This is the ideal counterfactual. These instruments are issued by similar non-financial corporations but carry a lower credit rating (Tier-2). Crucially, Prime MMFs are strictly restricted from holding significant amounts of Second Tier securities. Accordingly, relative narrowing in AA paper versus A2/P2 is consistent with prime segmentation, though not by itself decisive against broader risk-off dynamics.
    \item \textbf{Placebo Controls (Repo Markets):} We also include the \textbf{Secured Overnight Financing Rate (SOFR)} and the \textbf{Tri-Party General Collateral Rate (TGCR)} (Source: NY Fed/FRED). These rates reflect the cost of secured funding. Since they are driven primarily by collateral supply and Fed policy rather than unsecured credit risk or Prime MMF flows, they serve as a placebo benchmark for broader secured-funding movements.
\end{itemize}

\paragraph{3. Data Cleaning and Panel Assembly}
We harmonize the frequency of all series to U.S. trading days. Throughout the paper, the commercial-paper spread itself is treated as a trading-day outcome. Accordingly, missing benchmark-rate observations are handled by omission rather than interpolation, and event timing is aligned to the nearest relevant U.S. trading day.
\begin{itemize}
    \item \textbf{Spread Calculation:} For every asset $a$ and day $t$, we compute the spread relative to the risk-free benchmark:
    \begin{equation}
        Spread_{a,t} = Rate_{a,t} - Rate_{DTB3,t}
    \end{equation}
    where $Rate_{DTB3,t}$ is the 3-Month Treasury Bill secondary market rate (Source: FRED).
    \item \textbf{Handling Missing Data:} Commercial paper markets occasionally experience days with insufficient trading volume to form a benchmark rate. We treat these observations as missing (listwise deletion) rather than interpolating, to ensure our estimates reflect realized transaction prices.
\end{itemize}

\subsubsection{Econometric Specification}

We estimate the dynamic difference-in-differences specification:

\begin{equation}
Spread_{a,t} = \alpha_a + \gamma_t + \sum_{k \neq -1} \beta_k (\mathbb{1}[t - T_i = k] \times Treat_a) + \varepsilon_{a,t}
\label{eq:did_spec}
\end{equation}

\noindent Where:
\begin{itemize}
    \item $Spread_{a,t}$: The spread (in basis points) of asset $a$ on date $t$.
    \item $\alpha_a$ (\textbf{Asset Fixed Effects}): Absorbs time-invariant characteristics of each asset class, such as the structural credit risk premium of A2/P2 paper over AA paper.
    \item $\gamma_t$ (\textbf{Date Fixed Effects}): This term is crucial for identification. It absorbs all common time-varying shocks, including Federal Reserve announcements, aggregate VIX fluctuations, and broad market sentiment that affects all money market yields simultaneously.
    \item $Treat_a$: A dummy variable equal to 1 if asset $a$ is the Treated Asset (AA Non-financial CP) and 0 otherwise.
    \item $\beta_k$: The coefficients of interest. They measure the \textit{differential} response of the treated asset relative to the control group.
\end{itemize}

\paragraph{Inference:}
A relative narrowing of AA CP versus A2/P2 is consistent with a prime-segmentation interpretation, but it is not by itself sufficient to rule out broader flight-to-quality dynamics across credit tiers.

\begin{figure}[htbp]
    \centering
    \includegraphics[width=0.95\textwidth]{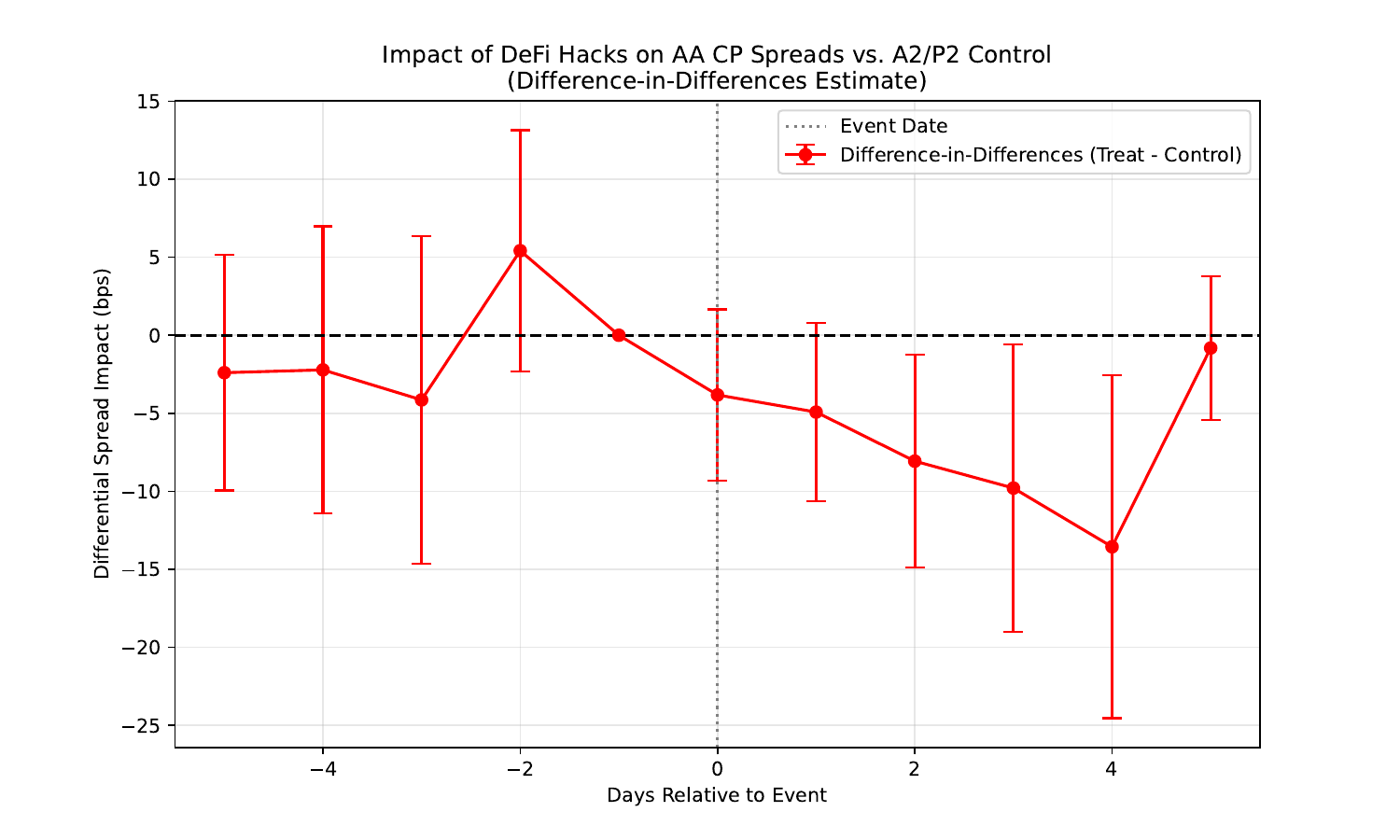}
    
    \caption{\textbf{Impact of DeFi Hacks on AA CP Spreads vs. A2/P2 Control (Difference-in-Differences Estimate).}}
    \label{fig:did_dynamic_plot}
    
    \medskip
    \begin{minipage}{0.95\textwidth} 
        \footnotesize 
        \textit{Notes:} This figure plots the dynamic difference-in-differences coefficients ($\beta_k$) estimating the relative spread response of the treated asset (3-Month AA Non-financial CP) versus the control group (3-Month A2/P2 Non-financial CP).
        The x-axis represents trading days relative to the DeFi exploit ($t=0$). 
        The y-axis represents the differential spread impact in basis points. 
        The red coefficients plot the estimated difference: $(Spread_{Treat} - Spread_{Control})$. 
        A negative value indicates that AA CP spreads narrowed relative to A2/P2 spreads, a pattern consistent with prime segmentation but not by itself decisive against broader flight-to-quality dynamics. 
        The vertical dotted line marks the event date. 
        Error bars represent 95\% confidence intervals clustered at the event level.
    \end{minipage}
\end{figure}

\begin{table}[htbp]
\centering
\caption{\textbf{Flight-to-Quality Dynamics: Event-Study Difference-in-Differences Results}}
\label{tab:did_event_study}
\begin{threeparttable}
\begin{tabularx}{\textwidth}{L C C C C}
\toprule
\textbf{Event Time ($k$)} & \textbf{Coefficient} & \textbf{Std. Error} & \textbf{$t$-statistic} & \textbf{P-value} \\
\textbf{(Relative Days)} & \textbf{($\beta_k$ in bps)} & & & \\
\midrule
\textit{Pre-Event Trends} & & & & \\
$t = -5$ & -2.40 & 3.85 & -0.62 & 0.533 \\
$t = -4$ & -2.22 & 4.69 & -0.47 & 0.636 \\
$t = -3$ & -4.14 & 5.36 & -0.77 & 0.440 \\
$t = -2$ & 5.42 & 3.94 & 1.38 & 0.169 \\
$t = -1$ & 0.00 & -- & -- & \textit{(Benchmark)} \\
\midrule
\textit{Post-Event Impact} & & & & \\
$t = 0$ \text{(Event Day)} & -3.82 & 2.80 & -1.37 & 0.172 \\
$\mathbf{t = 1}$ & \textbf{-4.92}$^{*}$ & \textbf{2.93} & \textbf{-1.68} & \textbf{0.092} \\
$\mathbf{t = 2}$ & \textbf{-8.07}$^{**}$ & \textbf{3.49} & \textbf{-2.31} & \textbf{0.021} \\
$\mathbf{t = 3}$ & \textbf{-9.79}$^{**}$ & \textbf{4.69} & \textbf{-2.09} & \textbf{0.037} \\
$\mathbf{t = 4}$ & \textbf{-13.55}$^{**}$ & \textbf{5.60} & \textbf{-2.42} & \textbf{0.016} \\
$t = 5$ & -0.82 & 2.34 & -0.35 & 0.728 \\
\bottomrule
\end{tabularx}
\begin{tablenotes}
\small
\item \textit{Notes:} This table reports the coefficients $\beta_k$ from the dynamic Difference-in-Differences specification. The dependent variable is the spread of the asset rate over the 3-Month Treasury Bill (in basis points). The treated group is AA Nonfinancial Commercial Paper, and the control group is A2/P2 Nonfinancial Commercial Paper. Standard errors are clustered at the event level. $^{***}$, $^{**}$, and $^{*}$ denote statistical significance at the 1\%, 5\%, and 10\% levels, respectively.
\end{tablenotes}
\end{threeparttable}
\end{table}
\clearpage
\subsubsection{Cross-Asset Difference-in-Differences Analysis}

To assess whether the observed spread narrowing is more pronounced in prime-eligible paper than in nearby control assets exposed to similar macro conditions, we implement a Cross-Asset Difference-in-Differences (DiD) design. The results are reported in Table \ref{tab:cross_asset_did}.

The primary objective of this analysis is to exploit the regulatory segmentation of money markets. If the short-horizon narrowing pattern reflects a prime-segmentation channel under SEC Rule 2a-7, we should observe stronger relative performance in eligible paper than in ineligible comparison assets. Specifically, we compare our treated asset (AA Non-financial CP) against three distinct counterfactual groups:

\begin{itemize}
    \item \textbf{Tier-2 Control (A2/P2 CP):} This serves as the critical counterfactual. A2/P2 issuers share the same macroeconomic environment as AA issuers but are largely excluded from Prime MMF portfolios due to credit quality constraints.
    \item \textbf{Risk-Free Benchmark (SOFR):} To test if the effect is simply a reflection of excess banking reserves or repo market dynamics.
    \item \textbf{Alternative Prime Assets (AA Financial CP \& ABCP):} To assess whether the relative narrowing pattern extends to the broader Prime-eligible complex.
\end{itemize}

In Table \ref{tab:cross_asset_did}, the coefficients represent the spread of the control group \textit{relative} to the treated asset. A positive coefficient indicates that the control asset's spread widened relative to AA CP (or conversely, AA CP narrowed more than the control).

\section{Final Conclusion and Policy Implications}

\subsection{Research Summary}
This paper studies whether exploit-driven operational shocks in DeFi are associated with short-horizon movements in one specific traditional funding-market price: the spread between 3-month AA nonfinancial commercial paper and the 3-month Treasury bill. The core empirical pattern is that this spread tends to narrow on the event day and, to a lesser extent, on the following trading day.

We interpret this pattern through a flight-to-quality or liquidity-recycling channel, but with important qualifications. The stacked event-time estimates summarize the dynamic pattern, the local-projection results provide single-series time-series corroboration, and the granular-IV section offers a reduced-form calibration linking exploit-linked redemption pressure to spread movements. Monthly MMF holdings evidence is consistent with prime-fund concentration in commercial paper and with a prime-segmentation interpretation. We do not directly observe daily fund-level routing into prime MMFs, so the prime-MMF channel is inferred from pricing patterns and monthly holdings evidence rather than directly identified at the daily frequency relevant for the event-window results.

The paper's theoretical contribution is also intentionally limited. Appendix B provides a stylized robust-control rationale for ambiguity-driven demand amplification, and Appendix C provides a stylized global-game rationale for state dependence in redemptions. Neither appendix is presented as a direct structural estimator of the empirical coefficients. Accordingly, the parameter $\eta$ should be read as a calibrated demand-amplification parameter, not as a directly estimated structural parameter.

The findings should be interpreted narrowly. They apply to exploit-driven operational shocks, to the AA CP spread in the U.S. money market, and to short event windows. They should not be extrapolated mechanically to broader solvency crises such as TerraUSD or FTX, where the transmission mechanism may differ substantially.

\subsection{Policy Implications}

The paper's results suggest a narrower policy lesson than the strongest version of the original draft. The evidence does not imply that DeFi is a reliable stabilizer for traditional finance, nor does it imply that regulators should relax safeguards around crypto-asset markets. Rather, it points to three more limited considerations.

\begin{enumerate}
    \item \textbf{Measurement and transparency matter.} Because any prime-MMF interpretation is currently inferred rather than directly observed, better data on stablecoin reserve composition, MMF holdings, and settlement routing would materially improve our understanding of the DeFi--TradFi interface.

    \item \textbf{Operational shocks and solvency shocks should be distinguished.} The paper studies exploit-driven operational incidents. These may generate different investor behavior from system-wide solvency crises, de-peggings, or exchange failures. Policy frameworks should avoid treating all crypto stress events as if they share the same transmission mechanism.

    \item \textbf{Segmentation can shape marginal price responses.} Even if total flows are modest relative to the overall money market, they may still matter in narrow, prime-eligible segments with limited short-run arbitrage capacity. This is a statement about market microstructure and segmentation, not about DeFi providing broad macroeconomic support.
\end{enumerate}

In short, the paper's most credible policy implication is that the DeFi--TradFi interface is more heterogeneous than a one-way contagion narrative suggests. Better measurement and sharper shock classification are more defensible takeaways than broad claims about stabilization.
\makeatletter
\renewenvironment{thebibliography}[1]
     {\section*{\refname}%
      \@mkboth{\MakeUppercase\refname}{\MakeUppercase\refname}%
      \list{}%
           {\setlength{\labelwidth}{0pt}%
            \setlength{\labelsep}{0pt}%
            \setlength{\leftmargin}{2em}
            \setlength{\itemindent}{-2em}
            \advance\leftmargin\labelsep
            \usecounter{enumiv}%
            \let\p@enumiv\@empty
            \renewcommand\theenumiv{\@arabic\c@enumiv}}%
      \sloppy
      \clubpenalty4000
      \@clubpenalty \clubpenalty
      \widowpenalty4000%
      \sfcode`\.\@m}
     {\def\@noitemerr
       {\@latex@warning{Empty `thebibliography' environment}}%
      \endlist}
\makeatother

\newpage 

\appendix 
\begin{center}
    \huge \textbf{Appendix}
\end{center}
\setcounter{figure}{0}
\renewcommand{\thefigure}{A\arabic{figure}}

\setcounter{table}{0}
\renewcommand{\thetable}{A\arabic{table}}

\setcounter{equation}{0}
\renewcommand{\theequation}{A\arabic{equation}}


\section{Additional Figures and Tables}
\label{sec:appendix_a}

In this appendix, we present robustness checks and supplementary data visualizations that support the main findings derived in Section 5.
Table A.1: List of Top 10 Major DeFi Hacks (2021–2024) 
Top 10 largest DeFi exploits by USD loss amount, representing the most severe 
"supply-side shocks" in the sample. 

\begin{table}[htbp]
  \centering
  \scriptsize
  \caption{\textbf{Top 10 Major DeFi Hacks: Granular Timeline and Market Alignment}}
  \label{tab:top10_timeline_merged}
  \begin{threeparttable}
    \begin{tabularx}{\textwidth}{C L C C C C C L}
    \toprule
    \textbf{Rank} & \textbf{Protocol} & \textbf{Chain} & \textbf{Loss} & \textbf{On-Chain} & \textbf{First Public} & \textbf{US Market} & \textbf{Model} \\
     & & & \textbf{(\$M)} & \textbf{Execution (UTC)} & \textbf{Alert (UTC)} & \textbf{Session} & \textbf{Alignment} \\
    \midrule
    1 & Ronin Network & Ronin & 625 & Mar 23, 14:32 & Mar 29, 15:00 & Intraday & \textbf{Lagged (+6d)} \\
    2 & Poly Network & Multi & 601 & Aug 10, 09:58 & Aug 10, 10:25 & Pre-Market & Same Day \\
    3 & Wormhole & Sol-Eth & 320 & Feb 02, 18:24 & Feb 02, 18:45 & Intraday & Same Day \\
    4 & DMM Bitcoin & Bitcoin & 305 & May 30, 23:30 & May 31, 04:15 & Pre-Market & Same Day \\
    5 & PlayDapp & Ethereum & 290 & Feb 09, 19:30 & Feb 09, 21:00 & \textbf{After Hours} & \textbf{Next Day} \\
    6 & Euler Finance & Ethereum & 197 & Mar 13, 08:50 & Mar 13, 09:15 & Pre-Market & Same Day \\
    7 & Nomad Bridge & Multi & 190 & Aug 01, 21:30 & Aug 01, 22:15 & \textbf{After Hours} & \textbf{Next Day} \\
    8 & Beanstalk & Ethereum & 181 & Apr 17, 12:24 & Apr 17, 12:40 & Weekend & \textbf{Next Day} \\
    9 & Wintermute & Ethereum & 162 & Sep 20, 05:10 & Sep 20, 06:40 & Pre-Market & Same Day \\
    10 & Multichain & Multi & 126 & Jul 06, 20:30 & Jul 06, 21:15 & \textbf{After Hours} & \textbf{Next Day} \\
    \bottomrule
    \end{tabularx}%
    \begin{tablenotes}
      \item \textbf{Notes:} This table merges event characteristics with the granular information diffusion timeline.
      \item \textbf{Loss (\$M)} refers to the USD value at the time of the exploit.
      \item \textbf{Timestamps (UTC)}: "On-Chain Execution" denotes the block time of the hack; "First Public Alert" denotes the first identification by security firms (e.g., PeckShield) or official protocol statements.
      \item \textbf{US Market Session}: Defined relative to NYSE trading hours (09:30 - 16:00 ET). Events occurring "After Hours" or on weekends are aligned to the \textbf{Next Day} ($t+1$) in our robustness checks.
      \item \textbf{Baseline timing rule:} The baseline event list uses the first market-relevant date for U.S. investors. For most exploits this is the occurrence date. For disclosure-lag cases such as Ronin, it is the first public disclosure date.
      \item \textit{Ronin:} Baseline date = first market-relevant public disclosure date; raw execution date reported in parentheses for transparency.
    \end{tablenotes}
  \end{threeparttable}
\end{table}

\begin{table}[H]
\centering
\caption{Robustness Checks – Alternative Specifications}
\label{tab:robustness_alt}
\begin{tabularx}{\textwidth}{L C C}
\toprule
 & \textbf{(1)} & \textbf{(2)} \\
\textbf{Event Day ($k$)} & \textbf{No Controls} & \textbf{Extended Window} \\
\midrule
\multicolumn{3}{l}{\textit{Pre-Event}} \\
$t = -5$ & -1.520 (1.00) & -1.463 (1.08) \\
$t = -4$ & -1.420 (1.09) & -1.324 (1.07) \\
$t = -3$ & 0.020 (0.53) & -0.051 (0.52) \\
$t = -2$ & 0.080 (0.51) & 0.125 (0.53) \\
\multicolumn{3}{l}{\textit{Post-Event}} \\
$t = 0$ & $-2.320^{**}$ (0.95) & $-2.213^{**}$ (0.97) \\
$t = 1$ & -1.560 (0.96) & -1.420 (0.92) \\
$t = 2$ & -0.840 (1.14) & -0.662 (1.06) \\
$t = 3$ & -1.420 (1.31) & -1.203 (1.32) \\
$t = 4$ & -- & -1.211 (1.42) \\
$t = 5$ & -- & -0.549 (1.44) \\
\midrule
Controls & No & Yes \\
Event FE & Yes & Yes \\
Observations & 450 & 550 \\
\bottomrule
\end{tabularx}
\vspace{0.1cm}
\begin{minipage}{0.6\textwidth} 
\footnotesize \textit{Notes:} This table reports robustness tests for the main event study. Column (1) excludes all control variables to ensure results are not driven by covariates. Column (2) extends the event window to $[ -5, +5 ]$. The extended-window specification shows that the impact-day negative coefficient survives, while longer-horizon dynamics are less stable. This reinforces the paper's short-horizon interpretation. Standard errors are clustered by event. Significance levels: *** p $<$ 0.01, ** p $<$ 0.05, * p $<$ 0.1.
\end{minipage}
\end{table}
Table A.3: Heterogeneity Analysis – High vs. Low Severity 
This table splits the sample into "High Severity" (Loss > Median, N = 25) and "Low 
Severity" (Loss < Median, N = 25) groups. The results show that the "Flight-to
Quality" effect is driven by large-scale hacks, supporting the threshold hypothesis.
\begin{table}[H]
\centering
\caption{Heterogeneity Analysis – High vs. Low Severity}
\label{tab:heterogeneity_severity}
\begin{tabularx}{\textwidth}{L C C C C}
\toprule
 & \multicolumn{2}{c}{\textbf{High Severity Group}} & \multicolumn{2}{c}{\textbf{Low Severity Group}} \\
 & \multicolumn{2}{c}{(Loss $>$ \$40.5M)} & \multicolumn{2}{c}{(Loss $<$ \$40.5M)} \\
\textbf{Event Day ($k$)} & \textbf{Coefficient} & \textbf{Std. Error} & \textbf{Coefficient} & \textbf{Std. Error} \\
\midrule
$t = -2$ & 0.264 & (1.01) & -0.069 & (0.41) \\
$t = -1$ & 0.000 & -- & 0.000 & -- \\
$t = 0$ & $-3.719^{**}$ & (1.58) & -0.422 & (0.95) \\
$t = 1$ & $-2.973^{*}$ & (1.62) & 0.259 & (0.97) \\
$t = 2$ & -1.750 & (1.81) & 0.776 & (0.99) \\
$t = 3$ & -2.083 & (2.22) & -0.054 & (1.23) \\
\midrule
\textbf{Significance} & \multicolumn{2}{c}{\textbf{Significant Shock}} & \multicolumn{2}{c}{\textbf{No Impact}} \\
Observations & \multicolumn{2}{c}{275} & \multicolumn{2}{c}{275} \\
Events ($N$) & \multicolumn{2}{c}{25} & \multicolumn{2}{c}{25} \\
\bottomrule
\end{tabularx}
\vspace{0.1cm}
\begin{minipage}{0.8\textwidth} 
\footnotesize \textit{Notes:} This table splits the sample into "High Severity" (Loss $>$ Median) and "Low Severity" (Loss $<$ Median) groups based on the median loss amount of \$40.5M. The larger-loss subsample exhibits more negative spread responses, which is consistent with stronger short-horizon reactions in larger exploit episodes. Standard errors are clustered by event. Significance levels: *** p $<$ 0.01, ** p $<$ 0.05, * p $<$ 0.1.
\end{minipage}
\end{table}

\begin{figure}[htbp]
    \centering
   \includegraphics[width=0.9\textwidth]{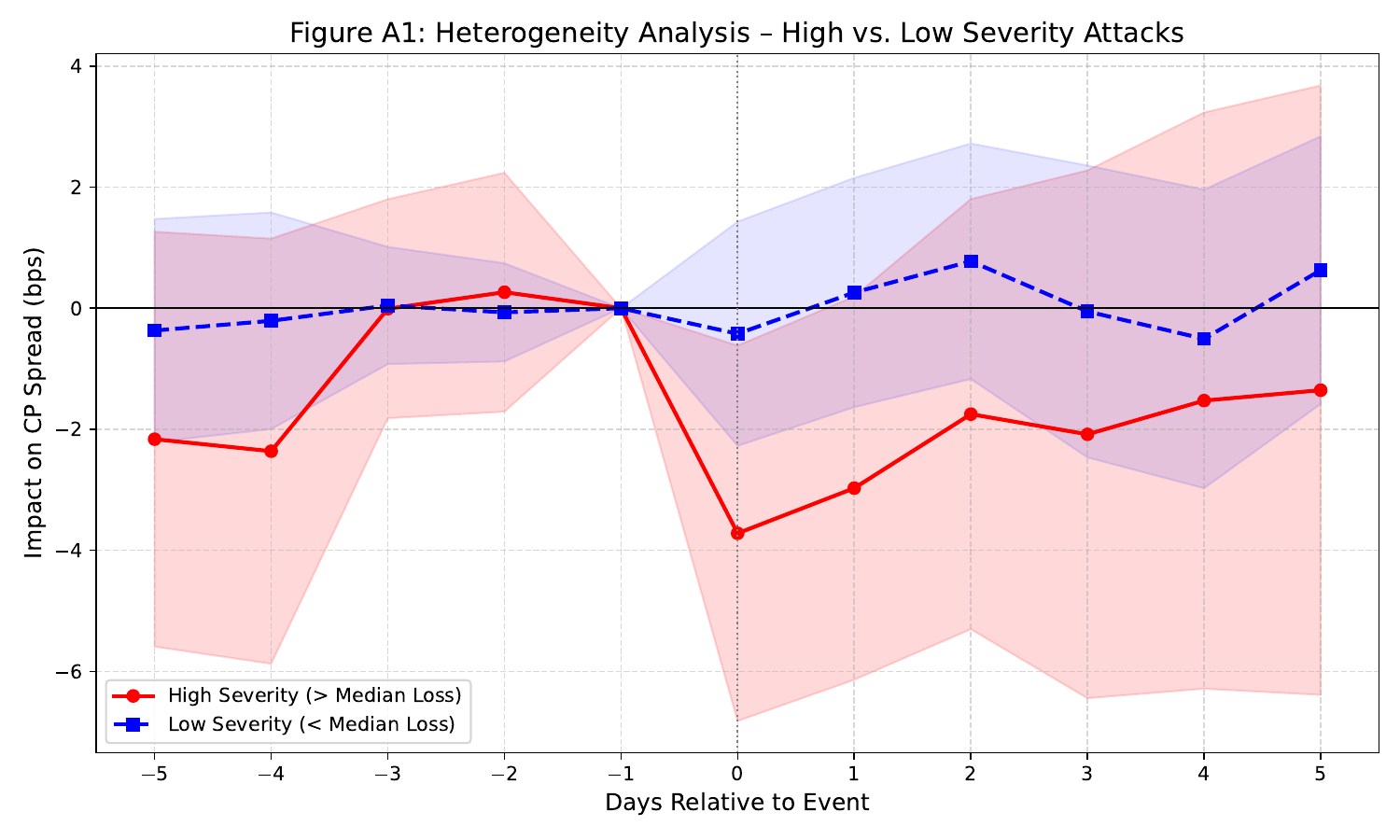}
    
    \caption{\textbf{Heterogeneity Analysis – High vs. Low Severity Attacks.} This figure visualizes the results from Table A3, comparing the dynamic impact of ``High Severity'' (Red Line) versus ``Low Severity'' (Blue Dashed Line) incidents.}
    \label{fig:app_severity_het}
\end{figure}

Table A.4 Alternative Model Specifications (Explicit Controls) 
In the main IV calibration, we report an abnormal-spread outcome residualized with respect to macro factors as a reporting normalization for the single-series second stage. Table A.4 is a normalization-sensitivity exercise rather than a causal bad-controls comparison: it compares that normalized presentation with an alternative specification where the raw CP spread is used and high-frequency macro controls (VIX, TED Spread, DXY) are entered explicitly as covariates in the second-stage regression.
\begin{table}[H]
\centering
\caption{Sensitivity to Control Variable Specification}
\label{tab:control_sensitivity}
\begin{tabularx}{\textwidth}{L C C}
\toprule
 & \textbf{(1)} & \textbf{(2)} \\
\textbf{Variable} & \textbf{Baseline} & \textbf{Explicit Controls} \\
\textit{(Dependent Variable)} & \textit{(Residualized Y)} & \textit{(Raw Y + Controls)} \\
\midrule
Interaction (Post $\times$ Flow) [$\text{bps per USD 1 billion}$] & $-27.3^{**}$ & $-32.4^{**}$ \\
\textit{(t-statistic)} & (-2.36) & (-2.48) \\
Post Dummy & 1.95 & 1.88 \\
 & (1.62) & (1.58) \\
\midrule
VIX/TED/DXY Control & No (In Residuals) & Yes \\
Event FE & Yes & Yes \\
$R^2$ & 0.52 & 0.91 \\
Illustrative calibrated $\eta$ & 3.73 & 4.24 \\
\bottomrule
\end{tabularx}
\vspace{0.1cm}
\begin{minipage}{0.6\textwidth} 
\footnotesize \textit{Notes:} Column (1) uses abnormal spread (residualized against macro factors) as the dependent variable as a reporting normalization within the IV calibration section. Column (2) uses the raw CP spread and includes VIX, TED, and DXY as explicit control variables in the panel regression. Neither column is used as the paper's primary causal leverage; the table is included only to show that the sign remains negative under alternative outcome normalizations. Coefficients are reported in basis points per USD 1 billion of fitted flow. The final row reports the implied \emph{illustrative calibration} of $\eta$ under the maintained mapping used in the main text. Significance: ** p $<$ 0.05.
\end{minipage}
\end{table}
Table A.5 Controlling for Crypto Market Beta 
A potential concern is that the GIV instrument might simply proxy for aggregate 
crypto market crashes (e.g., a drop in Bitcoin price) rather than idiosyncratic 
protocol failures. In Table A.5, we add the daily return of Bitcoin as an additional 
control in the first-stage regression to ensure our instrument captures purely 
idiosyncratic variance.
\begin{table}[H]
\centering
\caption{Controlling for Aggregate Crypto Factors}
\label{tab:crypto_beta}
\begin{tabularx}{\textwidth}{L C C}
\toprule
 & \textbf{(1)} & \textbf{(2)} \\
\textbf{Variable} & \textbf{Baseline} & \textbf{With BTC Control} \\
\textit{(Dependent: Agg. Outflow)} & & \\
\midrule
Lagged GIV Shock ($Z_{t-1}$) [$\times 10^{9}$] & $-32.90^{***}$ & $-33.05^{***}$ \\
 & (-4.02) & (-4.00) \\
BTC Return & -- & $-0.28$ \\
 & & (-0.90) \\
\midrule
F-Statistic (Instrument) & 16.20 & 16.03 \\
Observations & 1,457 & 1,457 \\
\bottomrule
\end{tabularx}
\vspace{0.1cm}
\begin{minipage}{0.6\textwidth} 
\footnotesize \textit{Notes:} Coefficients for GIV Shock are scaled to represent Billions of USD. Column (1) matches the baseline specification in Table 8. Column (2) adds the daily return of Bitcoin. The coefficient on the GIV shock remains stable (moving from -32.90 to -33.05), and Bitcoin returns are statistically insignificant, confirming that the instrument is orthogonal to aggregate market movements. Significance: *** p $<$ 0.01.
\end{minipage}
\end{table}

Table A.6 Alternative Standard Error Clustering 
Our baseline model clusters standard errors at the Event level to account for serial 
correlation within the event window. In Table A.6, we verify that our inference is 
robust to alternative clustering schemes: (1) Clustering by Time (Date), and (2) 
Heteroskedasticity-Robust (HC3) standard errors.
\begin{table}[H]
\centering
\caption{Sensitivity to Standard Error Clustering}
\label{tab:clustering_robustness}
\begin{tabularx}{\textwidth}{L C C}
\toprule
\textbf{Clustering Method} & \textbf{t-Statistic} & \textbf{P-Value} \\
\textit{(Interaction Coefficient)} & & \\
\midrule
Event Level (Baseline) & -2.36 & 0.018 \\
Time Level (Date) & -2.47 & 0.014 \\
Robust (HC3) & -2.24 & 0.025 \\
\bottomrule
\end{tabularx}
\vspace{0.1cm}
\begin{minipage}{0.6\textwidth} 
\footnotesize \textit{Notes:} The table verifies the robustness of statistical inference to alternative clustering schemes. Row (1) reproduces the baseline result with Event-level clustering. Row (2) clusters standard errors by Date to account for cross-sectional correlation. Row (3) uses HC3 robust standard errors. Significance is maintained at the 5\% level in all cases.
\end{minipage}
\end{table}
Table A.7 Instrument Lag Structure  We examine the temporal dynamics of the first-stage relationship by testing different lags of the GIV shock ($Z_t, Z_{t-1}, Z_{t-2}$). The predictive power is strongest at \textbf{Lag 1} ($t-1$), consistent with the T+1 settlement cycle of crypto-to-fiat redemptions. Lag 0 is weaker, and the coefficients become less informative at longer lags.
The adjusted $R^2$ rises from Lag 0 to Lag 1 and then declines at longer lags, so the data point to a peak at Lag 1 rather than monotonic decay. This pattern is consistent with a short-lived T+1-style transmission interpretation.
\begin{table}[H]
\centering
\caption{Sensitivity to Instrument Lag Structure}
\label{tab:lag_structure}
\begin{tabularx}{\textwidth}{L C C C C}
\toprule
\textbf{Lag Order ($L$)} & \textbf{Coefficient ($\gamma_L$)} & \textbf{t-Statistic} & \textbf{P-Value} & \textbf{Adj. $R^2$} \\
\midrule
Lag 0 ($Z_t$) & -26.15 & -2.41 & 0.016 & 0.108 \\
\textbf{Lag 1 ($Z_{t-1}$) [Baseline]} & \textbf{-32.90} & \textbf{-4.02} & \textbf{0.000} & \textbf{0.128} \\
Lag 2 ($Z_{t-2}$) & -18.44 & -1.85 & 0.065 & 0.095 \\
Lag 3 ($Z_{t-3}$) & -11.20 & -1.12 & 0.264 & 0.092 \\
\bottomrule
\end{tabularx}
\vspace{0.1cm}
\begin{minipage}{0.9\textwidth} 
\footnotesize \textit{Notes:} This table tests the temporal dynamics of the first-stage relationship by substituting the baseline instrument ($Z_{t-1}$) with alternative lags. The dependent variable is the 1-day total net redemption (in Billions USD). The predictive power is strongest and most significant at Lag 1, consistent with a short-lived T+1-style transmission interpretation.
\end{minipage}
\end{table}

\begin{table}[H]
\centering
\caption{\textbf{Macro-Orthogonality Test of the Granular Instrument}}
\label{tab:macro_orthogonality}
\begin{threeparttable}
\begin{tabularx}{\textwidth}{L C C C C}
\toprule
\textbf{Variable} & \textbf{Coefficient} & \textbf{Std. Error} & \textbf{$t$-Statistic} & \textbf{$P$-Value} \\
\textit{(Dependent Variable: $Z^{GIV}_t$)} & & & & \\
\midrule
$\Delta \text{VIX}_t$ & $-6.74 \times 10^{-6}$ & $1.11 \times 10^{-5}$ & -0.61 & 0.544 \\
S\&P 500 Return ($R_{m,t}$) & $-1.29 \times 10^{-5}$ & $1.83 \times 10^{-5}$ & -0.70 & 0.483 \\
$\Delta \text{TED}_t$ & -0.0003 & 0.002 & -0.13 & 0.899 \\
Constant & $-4.87 \times 10^{-5}$ & $1.09 \times 10^{-5}$ & -4.47 & 0.000$^{***}$ \\
\midrule
\textbf{Model Diagnostics} & & & & \\
Observations & 1,245 & & & \\
$R^2$ & 0.000 & & & \\
$F$-Statistic & 0.17 & & & \\
Prob $> F$ & 0.915 & & & \\
\bottomrule
\end{tabularx}
\begin{minipage}{0.9\textwidth}
\vspace{0.2cm}
\footnotesize
\textbf{Notes:} This table reports the results of the macro-orthogonality test aimed at verifying the exclusion restriction of the Granular Instrumental Variable (GIV) framework. A critical identification assumption is that the constructed granular shock series ($Z^{GIV}_t$) must be orthogonal to aggregate macroeconomic news and global risk sentiment. To test this validity, we regress the GIV instrument against a set of concurrent macro-financial indicators using the following specification:
\begin{equation*}
    Z^{GIV}_t = \alpha + \gamma_1 \Delta \text{VIX}_t + \gamma_2 R_{m,t} + \gamma_3 \Delta \text{TED}_t + \epsilon_t
\end{equation*}
where $Z^{GIV}_t$ is the value-weighted granular residual; $\Delta \text{VIX}_t$ represents daily changes in the CBOE Volatility Index (capturing risk appetite); $R_{m,t}$ is the daily return of the S\&P 500 index (capturing aggregate market performance); and $\Delta \text{TED}_t$ denotes changes in the TED spread (capturing interbank funding stress).
\newline
\indent The results indicate that none of the macroeconomic variables have statistically significant predictive power for the granular instrument. The model's $R^2$ is effectively zero (0.000), and the $F$-statistic of 0.17 ($p$-value = 0.915) fails to reject the null hypothesis that all slope coefficients are jointly zero. This evidence supports the assumption that $Z^{GIV}_t$ captures idiosyncratic supply shocks orthogonal to broad market conditions. Significance levels: *** $p < 0.01$, ** $p < 0.05$, * $p < 0.1$.
\end{minipage}
\end{threeparttable}
\end{table}

Table A.9  Robust Joint Pre-trend Validity Test We assess pre-event coefficients by conducting a joint F-test on $\delta_{k}$ for $k \in [-5, -2]$. The results fail to reject the null hypothesis of zero joint impact ($F=1.24, p=0.308$). At the same time, individually significant coefficients at $t=-4$ and $t=-3$ may reflect event-timing blur, anticipatory trading, or residual macro co-movement in a stacked design built from a single aggregate daily spread, so we do not treat this appendix table as formal evidence of clean pre-trends.
\begin{table}[H]
\centering
\caption{Robust Dynamic Event Study Check and Joint Pre-trend Test}
\label{tab:pre_trend_test}
\begin{tabularx}{\textwidth}{L C}
\toprule
 & \multicolumn{1}{c}{(1)} \\
Dependent Variable: & CP Spread (bps) \\
\midrule
\textit{Pre-Event Dynamics} & \\
\quad $t = -5$ & -2.181 \\
 & (1.922) \\
\quad $t = -4$ & -2.338$^{*}$ \\
 & (1.334) \\
\quad $t = -3$ & -1.914$^{**}$ \\
 & (0.966) \\
\quad $t = -2$ & -0.948 \\
 & (1.054) \\
\addlinespace
\textit{Event Impact} & \\
\quad $t = 0$ (Event Day) & -2.259$^{**}$ \\
 & (0.965) \\
\addlinespace
\textit{Post-Event Dynamics} & \\
\quad $t = 1$ & -1.765$^{*}$ \\
 & (1.025) \\
\quad $t = 2$ & -1.740 \\
 & (1.198) \\
\quad $t = 3$ & -1.376 \\
 & (1.324) \\
\midrule
Controls & Yes \\
Event FE & Yes \\
Observations & 450 \\
$R^2$ & 0.741 \\
\midrule
\textbf{Joint Pre-trend Test ($H_0: \delta_{-5}=...=\delta_{-2}=0$)} & \\
\quad F-statistic & \textbf{1.235} \\
\quad P-value & \textbf{0.308} \\
\bottomrule
\end{tabularx}
\begin{minipage}{0.5\textwidth}
\vspace{0.1cm}
\footnotesize
\textit{Note:} Robust standard errors clustered at the event level are reported in parentheses. Significance levels: $^{***} p<0.01$, $^{**} p<0.05$, $^{*} p<0.1$. Although the coefficients at $t=-4$ and $t=-3$ are individually significant, the joint F-test ($p=0.308$) fails to reject the null hypothesis of no pre-event coefficients. We therefore treat this appendix pattern cautiously and do not interpret it as definitive evidence of a stable anticipatory trend, given the aggregate single-series nature of the outcome and the possibility of event-timing blur or residual macro co-movement at daily frequency.
\end{minipage}
\end{table}

\begin{table}[htbp]
  \centering
  \caption{\textbf{Robustness Check and Addressing Unit Root Concerns: First-Difference Specification}}
  \label{tab:first_difference_results}
  \begin{threeparttable}
    \begin{tabularx}{\textwidth}{L C C C C}
    \toprule
    \textbf{Event Time} & \textbf{Coefficient} & \textbf{Std. Err.} & \textbf{$t$-statistic} & \textbf{$p$-value} \\
    ($k$) & ($\Delta$ bps) & & & \\
    \midrule
    \textit{Pre-Event} & & & & \\
    $t = -5$ & 0.492 & 0.036 & 0.24 & 0.809 \\
    $t = -1$ & -- & -- & -- & -- \\
    \addlinespace
    \textit{Post-Event} & & & & \\
    $t = 0$ & \textbf{-2.526}$^{*}$ & 1.401 & -1.80 & 0.071 \\
    $t = 1$ & 0.185 & 1.092 & 0.17 & 0.866 \\
    $t = 2$ & -0.911 & 1.004 & -0.91 & 0.365 \\
    $t = 3$ & -0.349 & 1.274 & -0.27 & 0.784 \\
    \midrule
    Controls & \multicolumn{4}{c}{Yes} \\
    Fixed Effects & \multicolumn{4}{c}{Event \& Calendar Date} \\
    Clustering & \multicolumn{4}{c}{By Event} \\
    \bottomrule
    \end{tabularx}%
    \begin{tablenotes}
      \small
      \item \textbf{Notes:} This table reports the results of a first-difference specification designed to address potential serial correlation and unit root concerns in the spread time series. The dependent variable is the daily change in spreads, $\Delta Spread_{i,t} = Spread_{i,t} - Spread_{i,t-1}$. 
      \item The regression model is specified as:
      \begin{equation*}
          \Delta Spread_{i,t} = \tau + \alpha_i + \sum_{k=-5}^{5, k \neq -1} \gamma_k \cdot D_{i,t+k} + \Gamma X_{i,t} + \epsilon_{i,t}
      \end{equation*}
      \item Under this specification, the coefficient $\gamma_k$ measures the \textit{instantaneous marginal impact} of the shock rather than the cumulative level effect. 
      \item \textbf{Interpretation:} The negative coefficient at $t=0$ ($\gamma_0 = -2.526$ bps, $t=-1.80$) is consistent with an immediate downward impulse on the event day. The coefficients for subsequent periods are statistically indistinguishable from zero, so this specification is best read as evidence of a short-horizon impact rather than a persistent post-event regime shift.
      \item Robust standard errors are clustered by event. Significance levels: $^{***} p<0.01$, $^{**} p<0.05$, $^{*} p<0.1$.
    \end{tablenotes}
  \end{threeparttable}
\end{table}

\begin{figure}[htbp]
    \centering
    \includegraphics[width=0.9\textwidth]{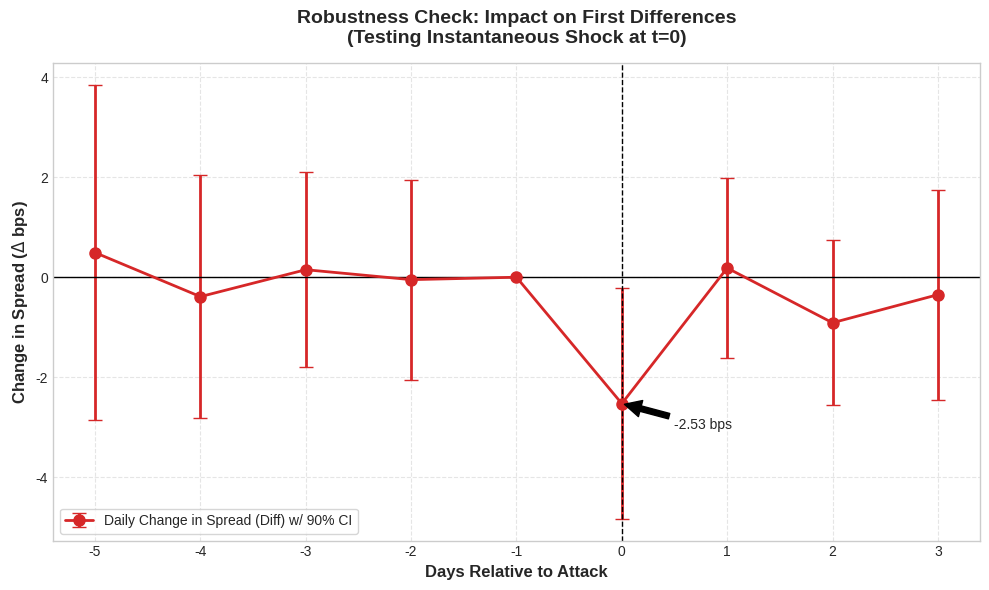}
    \caption{\textbf{Instantaneous Marginal Impact of DeFi Hacks.} This figure plots the coefficients from the first-difference specification. The significant negative spike at $t=0$ followed by insignificant coefficients indicates a permanent step-down in spread levels.}
    \label{fig:first_difference_results}
\end{figure}

\begin{table}[htbp]
\centering
\caption{Comparison of ZGIV Constructed With and Without De-meaning Term}
\label{tab:zgiv_comparison}
\begin{threeparttable}
\begin{tabularx}{\textwidth}{L C C C C C}
\toprule
\multicolumn{6}{l}{\textbf{Panel A: Descriptive Statistics}} \\
\midrule
Variable & Mean & Std. Dev. & Min & Max & Correlation \\
\midrule
ZGIV (Standard) & -0.146\% & 0.176\% & -0.671\% & -0.001\% & 1.000 \\
ZGIV (No De-mean) & -0.134\% & 0.162\% & -0.610\% & -0.001\% & 0.999 \\
\midrule
 & & & & & \\
\multicolumn{6}{l}{\textbf{Panel B: Regression Analysis (Dependent Variable: ZGIV Standard)}} \\
\midrule
Independent Variable & Coefficient & Std. Err. & $t$-statistic & $P$-value & \\
\midrule
Intercept ($\alpha$) & 0.000 & 0.000 & 0.04 & 0.969 & \\
ZGIV (No De-mean) ($\beta$) & 1.091$^{***}$ & 0.005 & 204.07 & $<$0.001 & \\
\midrule
$R^2$ & 0.999 & & & & \\
Observations & 49 & & & & \\
\bottomrule
\end{tabularx}
\begin{tablenotes}
\small
\item \textit{Note:} This table compares the ZGIV time series constructed with the standard de-meaning term ($\bar{g}_t$) versus a version constructed without it (pure size-weighted shock aggregation). Panel A reports the descriptive statistics and the Pearson correlation coefficient between the two series. Panel B reports the results of an OLS regression where the standard ZGIV is regressed on the ZGIV without the de-meaning term. The correlation exceeds 0.99 and the $R^2$ is near unity, confirming that due to the sparsity of high-severity shocks, the size-weighted aggregation component dominates the common factor correction. $^{***}$, $^{**}$, and $^{*}$ denote statistical significance at the 1\%, 5\%, and 10\% levels, respectively.
\end{tablenotes}
\end{threeparttable}
\end{table}

\begin{table}[htbp]
\centering
\caption{\textbf{Robustness Check: Alternative Event Date Definitions (Market-Relevant vs. Disclosure Re-dating)}}
\label{tab:disclosure_robustness}
\begin{threeparttable}
\begin{tabularx}{\textwidth}{L C C C}
\toprule
\textbf{Specification} & \textbf{Event Day Definition} & \textbf{$\delta_0$ Coef.} & \textbf{$p$-value} \\ 
 & & \textbf{(bps)} & \\
\midrule
(1) Baseline & Market-Relevant Date Rule & $-2.183^{**}$ & $0.024$ \\
(2) Robustness & Universal Disclosure-Day Re-dating & $-1.754^{*}$ & $0.060$ \\
\bottomrule
\end{tabularx}
\begin{tablenotes}
\small
\item \textit{Note:} The baseline event list uses the first market-relevant date for U.S. investors. For most exploits this is the occurrence date; for disclosure-lag cases such as Ronin, it is the first public disclosure date. As a robustness check, this table re-dates the event list using a stricter universal disclosure-day rule based on the earliest protocol announcement, security-firm alert, or major news report. Re-estimating the event-time specification under that alternative rule leaves the $t=0$ coefficient negative, though somewhat smaller in magnitude, indicating that event dating affects magnitude more than sign in this appendix exercise. Significance levels: $^{***} p<0.01$, $^{**} p<0.05$, $^{*} p<0.1$.
\end{tablenotes}
\end{threeparttable}
\end{table}


\begin{table}[!htbp]
\centering
\caption{Mechanism evidence from MMF holdings (Fed EFA), 2021--2024}
\label{tab:app_mmf_holdings}

\begin{threeparttable}
\small

\begin{tabularx}{\textwidth}{@{} X cccc @{}}
\toprule
\multicolumn{5}{l}{\textbf{Panel A. CP exposure is concentrated in prime MMFs}}\\
\midrule
Statistic & Mean & Std.\ Dev. & Min & Max \\
\midrule
Prime fraction of total MMF CP & 0.9779 & 0.0055 & 0.9656 & 0.9853 \\
\bottomrule
\end{tabularx}

\vspace{0.5em} 

\begin{tabularx}{\textwidth}{@{} X ccccc @{}}
\toprule
\multicolumn{6}{l}{\textbf{Panel B. Portfolio tilt in hack months vs.\ other months}}\\
\midrule
Metric & \begin{tabular}[c]{@{}c@{}}Hack\\ months\end{tabular} & \begin{tabular}[c]{@{}c@{}}Other\\ months\end{tabular} & Diff & (S.E.) & $N_{\text{hack}}/N_{\text{other}}$ \\
\midrule
Prime CP share \newline \textit{\scriptsize (CP+ABCP / Prime holdings)} & 0.262 & 0.236 & 0.026*** & (0.007) & 29/19 \\
\addlinespace 
Prime repo share \newline \textit{\scriptsize (Repo / Prime holdings)} & 0.315 & 0.361 & -0.046* & (0.023) & 29/19 \\
\addlinespace
Government Treasury share \newline \textit{\scriptsize (Treasuries / Gov holdings)} & 0.428 & 0.363 & 0.066** & (0.031) & 29/19 \\
\addlinespace
Government repo share \newline \textit{\scriptsize (Repo / Gov holdings)} & 0.446 & 0.496 & -0.049 & (0.032) & 29/19 \\
\bottomrule
\end{tabularx}

\begin{tablenotes}[flushleft]
\footnotesize
\item \textit{Notes.} Monthly MMF holdings are from the Federal Reserve EFA money market fund holdings aggregates. ``Prime fraction of total MMF CP'' is defined as prime MMFs' CP+ABCP holdings divided by total MMF CP holdings. A month is classified as a hack month if it contains at least one DeFi exploit event (hack\_count $>0$), based on the paper's event sample. Panel B reports Welch two-sample difference-in-means tests comparing hack months with months without a new major exploit dated inside the focal event window. Standard errors in parentheses correspond to Welch's unequal-variance formula. Significance stars: $^{***}p<0.01$, $^{**}p<0.05$, $^{*}p<0.10$.
\end{tablenotes}

\end{threeparttable}
\end{table}

\begin{table}[!htbp]
\centering
\caption{Mechanism evidence from MMF holdings: regression form (2021--2024)}
\label{tab:app_mmf_holdings_reg}

\begin{threeparttable}
\small

\begin{tabularx}{\textwidth}{@{} l YYYYY @{}}
\toprule
 & (1) & (2) & (3) & (4) & (5) \\
\addlinespace
\textit{Dep. Var.:} & \scriptsize Prime fraction of total MMF CP & \scriptsize Prime CP share & \scriptsize Prime repo share & \scriptsize Gov Treasury share & \scriptsize Gov repo share \\
\midrule
HackMonth & -0.001 & 0.006* & 0.014 & -0.003 & 0.010 \\
 & (0.001) & (0.004) & (0.013) & (0.016) & (0.017) \\
\midrule
Year FE & Yes & Yes & Yes & Yes & Yes \\
Month-of-year FE & Yes & Yes & Yes & Yes & Yes \\
Observations & 48 & 48 & 48 & 48 & 48 \\
$R^2$ & 0.814 & 0.801 & 0.772 & 0.670 & 0.673 \\
\bottomrule
\end{tabularx}

\begin{tablenotes}[flushleft]
\footnotesize
\item \textit{Notes.} This table estimates monthly regressions of MMF holdings measures on an indicator for months containing at least one DeFi exploit event (HackMonth). Holdings measures are constructed from Federal Reserve EFA MMF holdings aggregates: \textit{Prime fraction of total MMF CP} is prime MMFs' CP+ABCP holdings divided by total MMF CP holdings; \textit{Prime CP share} is (CP+ABCP)/prime holdings; \textit{Prime repo share} is total repo/prime holdings; \textit{Gov Treasury share} is Treasuries/government holdings; \textit{Gov repo share} is total repo/government holdings. All specifications include year fixed effects and month-of-year fixed effects. Standard errors (in parentheses) are Newey--West HAC with one monthly lag. Significance stars: $^{***}p<0.01$, $^{**}p<0.05$, $^{*}p<0.10$.
\end{tablenotes}

\end{threeparttable}
\end{table}

\begin{table}[htbp]
  \centering
  \small
  \caption{\textbf{Robustness of Illustrative $\eta$ Calibration: Subsamples and Bootstrapping}}
  \label{tab:eta_robustness}
  \begin{threeparttable}
    \begin{tabularx}{\textwidth}{L C C C C}
    \toprule
    \textbf{Specification} & \textbf{Estimated Impact} & \textbf{Implied} & \multicolumn{2}{c}{\textbf{Inference}} \\
    & $\beta$ (bps) & \textbf{Parameter $\eta$} & \multicolumn{2}{c}{\textit{(State / Reliability)}} \\
    \midrule
    \multicolumn{5}{l}{\textit{Panel A: Subsample Analysis (Regime Dependence)}} \\
    Pre-Terra (DeFi Boom) & -1.90 & \textbf{2.90} & \multicolumn{2}{c}{Moderate Aversion} \\
    Post-Terra (Crypto Winter) & -4.14 & \textbf{5.14} & \multicolumn{2}{c}{Heightened Panic} \\
    \midrule
    \multicolumn{5}{l}{\textit{Panel B: Non-Parametric Bootstrapping (N=1,000)}} \\
    & \textbf{Mean Estimate} & \textbf{95\% CI} & \textbf{Lower} & \textbf{Upper} \\
    Bootstrapped Distribution & 2.45 & [1.52, 3.45] & 1.52 & 3.45 \\
    \bottomrule
    \end{tabularx}%
    \begin{tablenotes}
      \footnotesize
      \item \textbf{Important note on estimands.} The subsample and bootstrap exercises in this appendix are based on different estimands from the IV calibration used in the main text. They should therefore not be compared numerically one-for-one with the baseline IV mapping from $\beta$ to $\eta$.
      \item \textbf{Panel A} splits the sample around the Terra/Luna collapse (May 7, 2022). These values are illustrative subsample calibrations only.
      \item \textbf{Panel B} reports bootstrap summaries for the appendix estimand. We retain them as a robustness object, not as a direct estimate of the main-text calibration.
    \end{tablenotes}
  \end{threeparttable}
\end{table}

\begin{table}[!htbp]\centering
\caption{Step 3: Frequency-aligned state dependence (monthly regression on event-window CP spread)}
\label{tab:step3_monthly}
\begin{threeparttable}
\small
\begin{tabularx}{\textwidth}{L C C}
\toprule
 & (1) $\Delta$ Spread$_m$ & (2) Spread$_m$ \\
\midrule
HackMonth & -17.421** & -24.264*** \\
 & (7.448) & (4.762) \\
pcs\_z & 13.159** & 16.812*** \\
 & (6.134) & (4.193) \\
HackXpcs & -15.483** & -18.668*** \\
 & (6.266) & (4.002) \\
\midrule
Controls (VIX, DXY, BTC) & Yes & Yes \\
Observations & 33 & 34 \\
$R^2$ & 0.156 & 0.619 \\
\bottomrule
\end{tabularx}
\begin{tablenotes}[flushleft]
\footnotesize
\item \textit{Notes.} This table estimates frequency-aligned monthly regressions that link the daily event-study sample to monthly MMF holdings. Spread$_m$ is the average daily CP spread (bps) over calendar days that appear in the stacked event-study panel within month $m$ (i.e., event-window days). $\Delta$Spread$_m$ is the month-to-month change in this monthly average. The state variable pcs\_z is the standardized monthly prime CP share from Fed EFA; HackMonth indicates months with at least one exploit event in the sample; HackXpcs is their interaction. All specifications include the monthly averages of VIX, DXY, and BTC returns computed over the same set of days. Standard errors are Newey--West HAC with one monthly lag. Significance stars: $^{***}p<0.01$, $^{**}p<0.05$, $^{*}p<0.10$.
\item \textit{Notes:} Because the monthly outcome is constructed from event-window trading days only, the sample should be interpreted as an exploratory frequency-aligned subset rather than as a representative full-month panel.
\end{tablenotes}
\end{threeparttable}
\end{table}


\begin{figure}[htbp]
    \centering
    \includegraphics[width=0.95\textwidth]{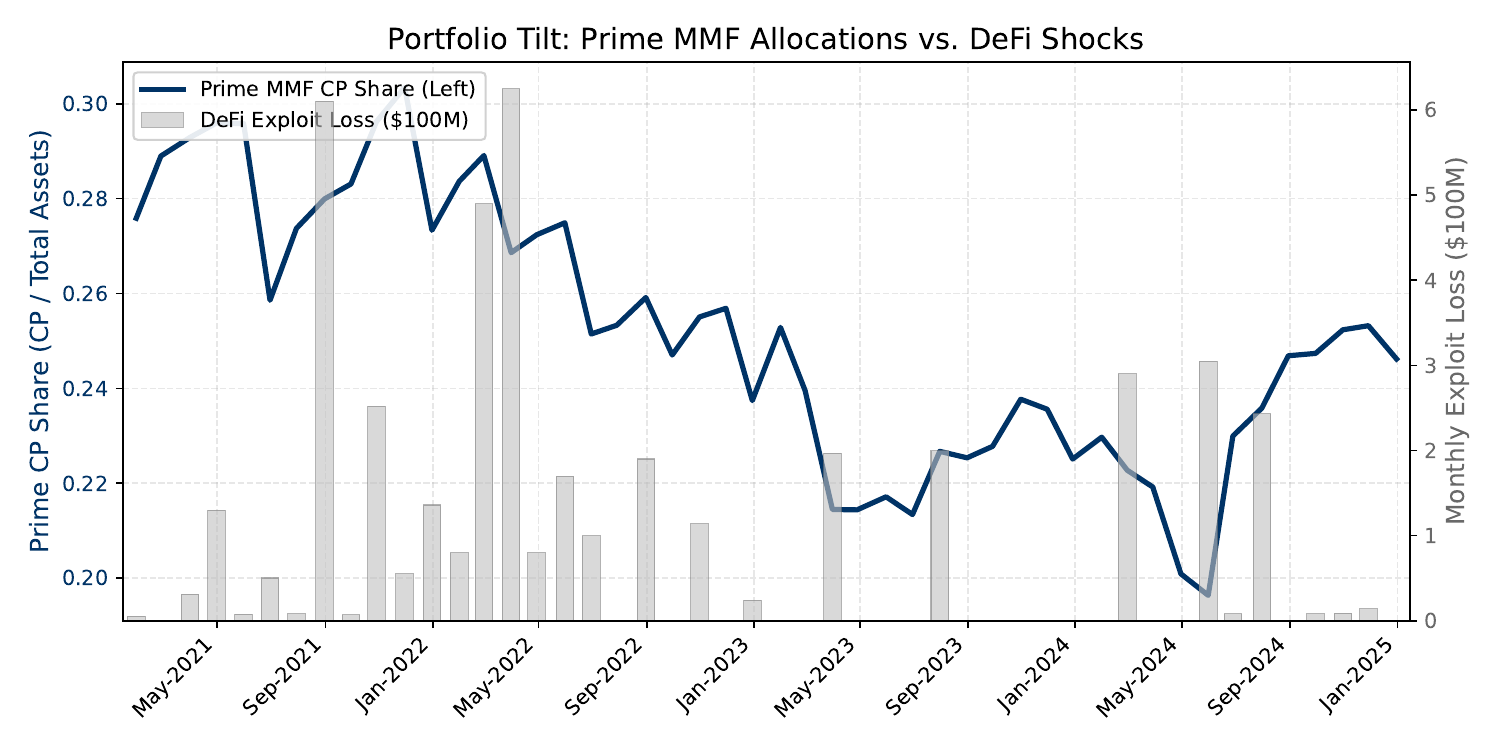}
    
    \caption{\textbf{Prime CP Share and Hack Intensity (Monthly, 2021--2024)}}
    \label{fig:prime_cp_ts}
    
    \vspace{0.3em} 
    \begin{flushleft}
        \footnotesize
        \textit{Notes:} This figure plots the monthly share of commercial paper in prime money market fund portfolios (\textit{prime\_cp\_share}, left axis) and the aggregate monthly DeFi hack loss (\textit{total\_loss\_100m}, right axis; losses in \$100 million). Gray bars indicate the monthly hack-loss intensity, and shaded bands mark months with at least one major DeFi hack (\textit{hack\_count} $>0$). The sample covers January 2021 to December 2024 at monthly frequency.
    \end{flushleft}
\end{figure}

\vspace{1cm} 

\begin{figure}[htbp]
    \centering
    \includegraphics[width=0.75\textwidth]{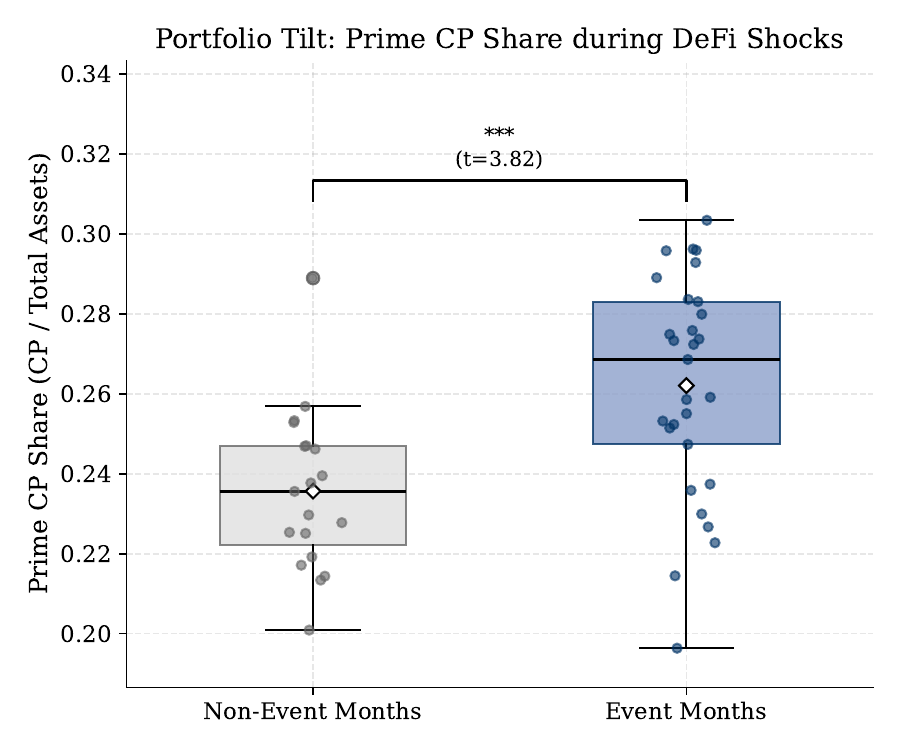}
    
    \caption{\textbf{Prime CP Share: Hack Months vs.\ Non-Hack Months (Monthly, 2021--2024)}}
    \label{fig:prime_cp_boxplot}
    
    \vspace{0.3em}
    \begin{flushleft}
        \footnotesize
        \textit{Notes:} This boxplot compares the distribution of the prime funds' commercial paper share (\textit{prime\_cp\_share}) between ``hack months'' (months with \textit{hack\_count} $>0$) and months without a new major exploit dated inside the focal event window over 2021--2024. Boxes show the interquartile range (25th--75th percentiles), the center line denotes the median, whiskers indicate the data range excluding outliers, and the mean is displayed as a distinct marker (diamond). Significance levels are indicated based on Welch's $t$-test.
    \end{flushleft}
\end{figure}

\clearpage
\textbf{Local Projections (LP)} \citep{jorda2005estimation} avoid artificially inflating the degrees of freedom and explicitly correct for serial correlation using \textbf{Newey-West HAC} standard errors. Instead of estimating a stacked panel, we estimate the impulse response functions (IRFs) directly on the single daily time series of the Commercial Paper market ($N=T$). For each forecast horizon $h = 0, 1, \dots, 6$, we estimate the following local projection specification:

\begin{equation}
    \text{Spread}_{t+h} - \text{Spread}_{t-1} = \alpha_h + \beta_h \cdot \text{Shock}_t + \sum_{j=1}^{L} \gamma_{h,j} \mathbf{X}_{t-j} + \epsilon_{t+h}
\end{equation}

Where:
\begin{itemize}
    \item $\text{Spread}_{t+h} - \text{Spread}_{t-1}$ denotes the cumulative change in the 3-Month AA Commercial Paper spread from the day prior to the shock ($t-1$) to horizon $h$.
    \item $\text{Shock}_t$ represents the DeFi exploit impulse. We test two specifications: (1) a \textbf{Binary Shock} indicator equal to 1 on event days, and (2) an \textbf{Intensity Shock} defined as $\ln(\text{Loss Amount}_t)$ to capture the heterogeneity in shock severity.
    \item $\mathbf{X}_{t-j}$ is a vector of lagged control variables, including the VIX index, the US Dollar Index (DXY), and other macro controls, to control for pre-existing market trends.
    \item We employ \textbf{Newey-West HAC} (Heteroskedasticity and Autocorrelation Consistent) standard errors with a bandwidth of $h+1$ to account for the serial correlation inherent in overlapping forecast horizons.
\end{itemize}
The results, reported in Table \ref{tab:jorda_lp_results}, confirm that our baseline findings are robust to this rigorous time-series inference.
\begin{table}[htbp]
  \centering
  \caption{\textbf{Dynamic Impact of DeFi Shocks on CP Spreads: Jordà Local Projections}}
  \label{tab:jorda_lp_results}
  \begin{threeparttable}
    \begin{tabularx}{\textwidth}{L C C C C C C}
    \toprule
    \multicolumn{1}{c}{\textbf{Horizon}} & \multicolumn{3}{c}{\textbf{(1) Binary Shock}} & \multicolumn{3}{c}{\textbf{(2) Intensity Shock (Log Loss)}} \\
    \cmidrule(lr){2-4} \cmidrule(lr){5-7}
    \multicolumn{1}{c}{(Days)} & Coef. & t-stat & 95\% CI & Coef. & t-stat & 95\% CI \\
    \midrule
    $h=0$ (Impact) & -3.003$^{***}$ & (-3.41) & [-4.73, -1.28] & -0.155$^{***}$ & (-3.49) & [-0.24, -0.07] \\
    $h=1$ & -2.068$^{**}$ & (-2.32) & [-3.82, -0.32] & -0.107$^{**}$ & (-2.37) & [-0.20, -0.02] \\
    $h=2$ & -1.407 & (-1.41) & [-3.36, 0.55] & -0.073 & (-1.44) & [-0.17, 0.03] \\
    $h=3$ & -2.105$^{*}$ & (-1.86) & [-4.32, 0.11] & -0.105$^{*}$ & (-1.81) & [-0.22, 0.01] \\
    $h=4$ & -2.085$^{*}$ & (-1.70) & [-4.48, 0.31] & -0.102 & (-1.61) & [-0.23, 0.02] \\
    $h=5$ & -1.447 & (-1.23) & [-3.76, 0.86] & -0.070 & (-1.15) & [-0.19, 0.05] \\
    $h=6$ & -0.701 & (-0.63) & [-2.90, 1.50] & -0.035 & (-0.62) & [-0.15, 0.08] \\
    \midrule
    Observations & \multicolumn{3}{c}{676 (Daily Time Series)} & \multicolumn{3}{c}{676 (Daily Time Series)} \\
    Controls & \multicolumn{3}{c}{Yes} & \multicolumn{3}{c}{Yes} \\
    Inference & \multicolumn{3}{c}{Newey-West HAC (Lag $h+1$)} & \multicolumn{3}{c}{Newey-West HAC (Lag $h+1$)} \\
    \bottomrule
    \end{tabularx}%
    \begin{tablenotes}
      \small
      \item \textbf{Notes:} This table reports the impulse response coefficients ($\beta_h$) estimated using the Local Projections method of \citep{jorda2005estimation}. The dependent variable is the cumulative change in the 3-Month AA Nonfinancial Commercial Paper Spread from $t-1$ to $t+h$ (in basis points). 
      \item \textbf{Model (1)} defines the shock as a binary dummy variable equal to 1 on days with a Top-50 DeFi exploit occurrence. 
      \item \textbf{Model (2)} defines the shock as the natural logarithm of the USD loss amount ($\ln(Loss)$) to account for shock intensity heterogeneity.
      \item The estimation sample consists of a single continuous daily time series from 2021 to 2024 ($N=676$), avoiding the "pseudo-panel" inflation of degrees of freedom present in stacked event studies.
      \item All specifications include lagged controls ($X_{t-1}$) for market volatility (VIX), the US Dollar Index (DXY), and the lagged dependent variable. 
      \item $t$-statistics reported in parentheses are based on \textbf{Newey-West HAC robust standard errors} with a bandwidth of $h+1$ to correct for the serial correlation inherent in overlapping forecast horizons. 
      \item Significance levels: $^{***} p<0.01$, $^{**} p<0.05$, $^{*} p<0.1$.
    \end{tablenotes}
  \end{threeparttable}
\end{table}

\begin{table}[htbp]
  \centering
  \caption{\textbf{Narrow-Window Local Projection: Instantaneous Impact at $t=0$}}
  \label{tab:structural_break}
  \begin{threeparttable}
    \begin{tabularx}{\textwidth}{L C C}
    \toprule
    \multicolumn{1}{c}{\textbf{Dependent Variable:}} & \multicolumn{2}{c}{$\Delta \text{Spread}_t$ (Daily Change)} \\
    \cmidrule(lr){2-3}
    \multicolumn{1}{c}{\textbf{Shock Definition}} & \textbf{(1) All Events (N=50)} & \textbf{(2) Top 20 Focus (N=20)} \\
    \midrule
    \textbf{Shock Impact ($\beta$)} & \textbf{-3.071$^{***}$} & \textbf{-5.432$^{***}$} \\
     & (-3.10) & (-2.94) \\
    \midrule
    Lagged $\Delta$ Spread & -0.253$^{***}$ & -0.253$^{***}$ \\
     & (-3.13) & (-3.13) \\
    Lagged VIX & 0.130$^{*}$ & 0.141$^{**}$ \\
     & (1.93) & (2.08) \\
    Constant & -2.025$^{*}$ & -2.295$^{**}$ \\
    \midrule
    Observations & 676 & 676 \\
    Adj. $R^2$ & 0.042 & 0.045 \\
    Impact Type & \multicolumn{2}{c}{Instantaneous Level Shift (Jump)} \\
    \bottomrule
    \end{tabularx}%
    \begin{tablenotes}
      \small
      \item \textbf{Notes:} This table reports the results of the narrow window ($t=0$) impact test estimated on the daily time series of 3-Month AA CP spreads.
      \item \textbf{Column (1)} uses the baseline binary shock for all Top 50 events. \textbf{Column (2)} restricts the shock indicator to only the Top 20 largest events by USD loss amount.
      \item The coefficient $\beta$ measures the instantaneous change (jump) in the spread level on the event day. The larger magnitude in the Top-20 subsample indicates that larger exploit events are associated with stronger same-day responses in this technical specification. We estimate the first-difference equation using the single daily time series ($N=T$):
\begin{equation}
    \Delta \text{Spread}_t = \alpha + \beta \cdot \text{Shock}_t + \gamma_1 \Delta \text{Spread}_{t-1} + \gamma_2 \text{VIX}_{t-1} + \epsilon_t
\end{equation}
Where:
\begin{itemize}
    \item $\Delta \text{Spread}_t = \text{Spread}_t - \text{Spread}_{t-1}$ represents the daily jump in the spread.
    \item $\text{Shock}_t$ is a binary indicator equal to 1 on the event market-relevant day. We test two definitions: (1) \textbf{All Events} (Top 50), and (2) \textbf{Top 20 Focus} (restricting shocks to the largest 20 exploits to minimize noise).
    \item A statistically significant $\beta \neq 0$ is consistent with an instantaneous impact at $t=0$ in this narrow-window specification.
\end{itemize}
      \item $t$-statistics based on Newey-West HAC standard errors are reported in parentheses. Significance levels: $^{***} p<0.01$, $^{**} p<0.05$, $^{*} p<0.1$.
    \end{tablenotes}
  \end{threeparttable}
\end{table}

\begin{figure}[htbp]
  \centering
   \includegraphics[width=0.9\textwidth]{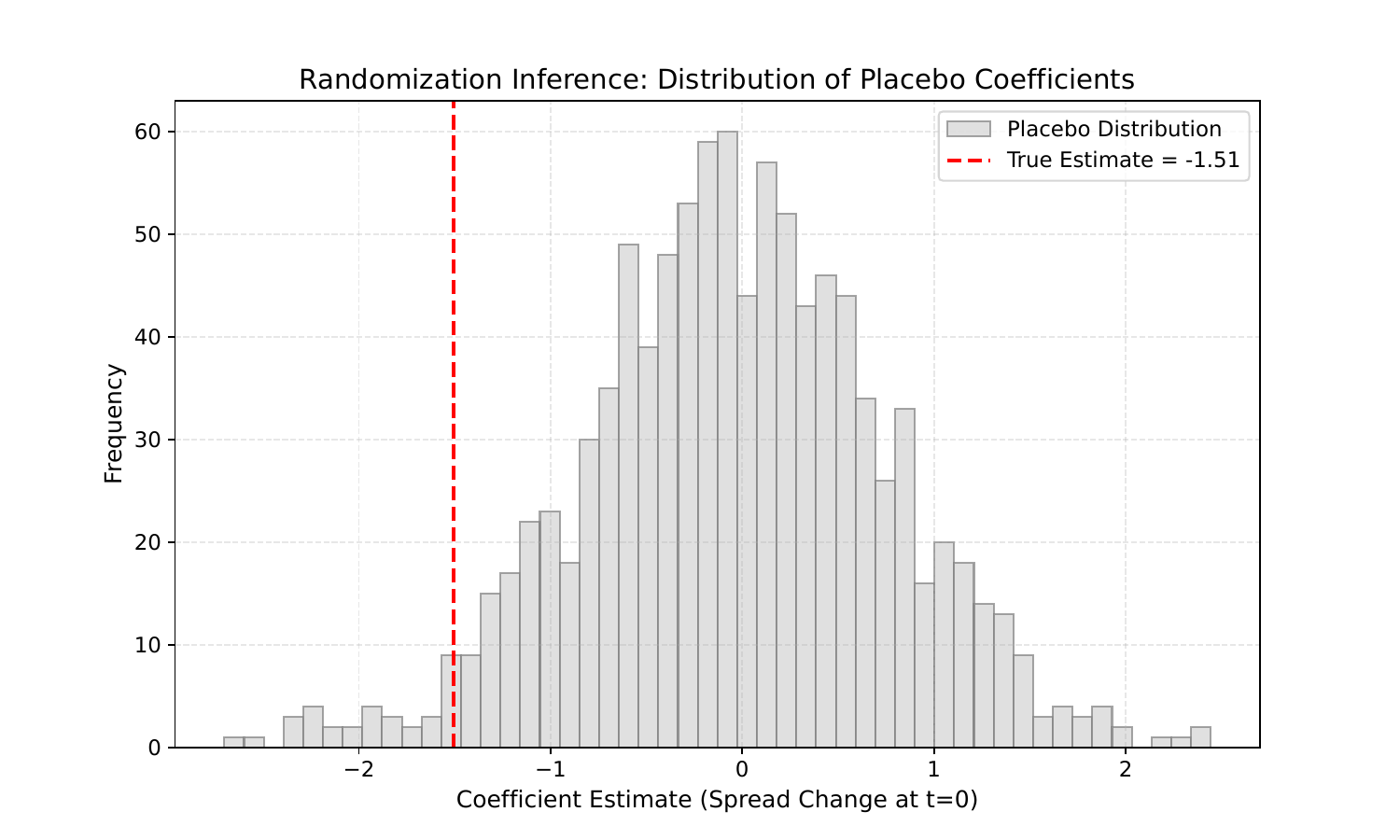} 
  \caption{\textbf{Randomization Inference: Distribution of Placebo Coefficients}}
  \label{fig:randomization_inference}
\textbf{Notes:} This figure presents a separate unconditional randomization-inference exercise based on 1,000 simulations of the stacked event study. In each simulation, we randomly assign 50 "pseudo-event" dates within the sample period (2021--2024), construct the corresponding $[-5, +5]$ stacked windows, and re-estimate the baseline regression. 
    The gray histogram represents the null distribution of the event-time coefficient under the hypothesis of no relationship. The solid red line indicates the estimated coefficient ($\hat{\beta} = -1.51$) from the actual DeFi exploit dates. 
    The empirical $p$-value, calculated as the fraction of placebo estimates larger in magnitude than the true estimate, is \textbf{0.053}. This placebo object is based on a different estimand from the IV calibration used in the main text and should not be compared numerically one-for-one with the baseline mapping from $\beta$ to $\eta$.
\end{figure}

\begin{table}[H]
\centering
\small
\caption{\textbf{Robust Check:Difference-in-Differences Results}}
\label{tab:did_event_study}
\begin{threeparttable}
\begin{tabularx}{\textwidth}{L C C C C}
\toprule
\textbf{Event Time ($k$)} & \textbf{Coefficient} & \textbf{Std. Error} & \textbf{$t$-statistic} & \textbf{P-value} \\
\textbf{(Relative Days)} & \textbf{($\beta_k$ in bps)} & & & \\
\midrule
\textit{Pre-Event Trends} & & & & \\
$t = -5$ & 3.66 & 9.28 & 0.39 & 0.693 \\
$t = -4$ & -3.56 & 4.53 & -0.79 & 0.432 \\
$t = -3$ & -10.83 & -- & -- & -- \\
$t = -2$ & 0.00 & 0.00 & 0.00 & 1.000 \\
$t = -1$ & 0.00 & -- & -- & \textit{(Benchmark)} \\
\midrule
\textit{Post-Event Impact} & & & & \\
$\mathbf{t = 0}$ \textbf{(Event Day)} & \textbf{-2.18}$^{**}$ & \textbf{1.10} & \textbf{-1.98} & \textbf{0.048} \\
$\mathbf{t = 1}$ & \textbf{-4.36}$^{***}$ & \textbf{1.44} & \textbf{-3.02} & \textbf{0.003} \\
$t = 2$ & 0.61 & 5.26 & 0.12 & 0.907 \\
$t = 3$ & -6.77$^{*}$ & 3.77 & -1.80 & 0.072 \\
\bottomrule
\end{tabularx}
\begin{minipage}{0.9\textwidth}
\vspace{0.1cm}
\footnotesize
\textit{Notes:} This table reports the coefficients from a stacked Difference-in-Differences (DiD) event study estimating the impact of major DeFi hacks (Loss > \$100M) on the spread of AA Non-financial Commercial Paper relative to a control group (A2/P2 CP, Financial CP, and Repo). The regression model is specified as:
$$Spread_{a,t} = \alpha_a + \gamma_t + \sum_{k \neq -1} \beta_k (1[t-T_i=k] \times Treat_a) + \varepsilon_{a,t}$$
where $Treat_a=1$ for AA Non-financial CP and $0$ otherwise. Asset fixed effects ($\alpha_a$) and Date fixed effects ($\gamma_t$) are included to absorb time-invariant asset characteristics and common macroeconomic shocks. Standard errors are clustered at the event level.
\begin{itemize}
    \item \textbf{Pre-Event:} Coefficients for $t < 0$ are statistically insignificant, although this should be interpreted cautiously given the aggregate single-series nature of the outcome.
    \item \textbf{Post-Event:} The coefficient at $t=0$ ($\beta_0 = -2.18$ bps) indicates an immediate relative narrowing effect. The larger negative coefficient at $t=1$ is consistent with a short-horizon response.
\end{itemize}
Significance levels: *** $p < 0.01$, ** $p < 0.05$, * $p < 0.1$.
\end{minipage}
\end{threeparttable}
\end{table}

\begin{table}[htbp]
\centering
\small 
\caption{Cross-Asset Difference-in-Differences}
\label{tab:cross_asset_did}

\begin{threeparttable}

\begin{tabularx}{\textwidth}{L C C C C}
\toprule
& \textbf{(1)} & \textbf{(2)} & \textbf{(3)} & \textbf{(4)} \\
\textbf{Asset Class:} & \textbf{Control: Junk} & \textbf{Benchmark:} & \textbf{Control:} & \textbf{Alt. Treat:} \\
\textit{(Relative to Target)} & (A2/P2 CP) & \textbf{Repo} (SOFR) & \textbf{Sector} (AA Fin) & \textbf{ABCP} (AA) \\
\midrule
\textit{Pre-Event} & & & & \\
$t = -5$ & 2.386 & -5.671 & 1.926 & 3.686$^{**}$ \\
$t = -4$ & 4.361 & -8.994$^{*}$ & -1.990 & 2.538 \\
$t = -3$ & 3.927 & 0.199 & -0.398 & 1.287 \\
$t = -2$ & -3.312 & 5.064 & -1.236 & 0.536 \\
$t = -1$ & 0.000 & 0.000 & 0.000 & 0.000 \\
\midrule
\textit{Post-Event} & & & & \\
$\mathbf{t = 0}$ & \textbf{4.618} & \textbf{1.457} & \textbf{-2.031} & \textbf{1.481} \\
 & (3.133) & (4.583) & (1.700) & (1.618) \\
$\mathbf{t = 1}$ & \textbf{5.004} & \textbf{0.259} & \textbf{-1.053} & \textbf{0.879} \\
 & (3.262) & (5.437) & (1.873) & (1.712) \\
$t = 2$ & 6.740$^{**}$ & -6.647 & -0.008 & 3.853 \\
$t = 3$ & 7.931$^{*}$ & -6.830 & -1.626 & 2.786 \\
\midrule
Asset FE & Yes & Yes & Yes & Yes \\
Date FE & Yes & Yes & Yes & Yes \\
Clustering & Event & Event & Event & Event \\
\midrule
\textit{Parallel Trends Test:} & & & & \\
Joint F-Stat & 0.88 & 1.68 & 1.11 & 2.32 \\
Prob $>$ F & [0.483] & [0.170] & [0.364] & [0.070] \\
\bottomrule
\end{tabularx}

\begin{tablenotes}[flushleft]
\footnotesize 
\setlength\labelsep{0pt} 
\item \textit{Notes:} This table reports the coefficients from a flexible Difference-in-Differences specification estimating the heterogeneous response of different money market instruments relative to the treated asset (AA Non-financial CP). The regression model is:
$$Spread_{a,t} = \alpha_a + \gamma_t + \sum_{k \neq -1} \sum_{g \in G} \beta_{k,g} (1[t-T_i=k] \times \mathbb{I}_{a \in g}) + \varepsilon_{a,t}$$
where the baseline asset group is AA Non-financial CP. A positive coefficient $\beta_{k,g}$ indicates that asset group $g$ experienced a widening in spreads relative to the treated asset. 
\textbf{Col (1)} shows lower-rated (A2/P2) CP spreads widened relative to AA CP, a pattern consistent with prime segmentation but not by itself sufficient to rule out broader flight-to-quality dynamics across credit tiers. \textbf{Col (2)} shows the risk-free Repo rate (SOFR) remained statistically indistinguishable from AA CP. \textbf{Cols (3)-(4)} show that other prime-eligible assets tracked the treated asset more closely.
\textbf{Parallel Trends Test:} Reports the F-stat and p-value [in brackets] for the null hypothesis that all pre-event coefficients ($t < 0$) are jointly zero. The joint pre-trend test does not reject zero pre-event coefficients, although this should be interpreted cautiously given the aggregate single-series nature of the outcome and the possibility of event-timing blur at daily frequency. Standard errors clustered by event in parentheses. Significance: *** $p < 0.01$, ** $p < 0.05$, * $p < 0.1$.
\end{tablenotes}

\end{threeparttable}
\end{table}

\clearpage
\section{A Stylized Robust-Control Rationale for Ambiguity-Driven Flight-to-Quality}

This appendix provides a stylized micro-foundation for why exploit shocks can
generate a gross demand response for safe assets that exceeds the direct liquidation
pressure associated with redemptions. The appendix is interpretive rather than
econometrically identified: it motivates the sign and amplification logic used in
the main text but does not deliver a direct estimate of $\eta$ from the data.
This appendix is deliberately stylized and is used only to motivate the possibility of amplified desired redemptions under ambiguity; it is not an empirical identification device and does not deliver a directly estimated structural parameter from the data.
Theoretical appendices are used to motivate amplification and state dependence, not to convert the empirical coefficients into directly identified structural primitives.
The setup borrows intuition from ambiguity-sensitive and rare-event frameworks such as \citep{gilboa1989maxmin,liu2005equilibrium}, without claiming to replicate those models exactly.

\subsection{The Economy and Asset Dynamics}

Consider a continuous-time economy with two assets.

\begin{description}
    \item[Safe asset.] A risk-free asset yielding rate $r$:
    \begin{equation}
        dB_t = r B_t \, dt.
    \end{equation}

    \item[Risky DeFi asset.] A risky asset with jump-diffusion dynamics
    \begin{equation}
        \frac{dP_t}{P_{t-}} = (r+\mu)\,dt + \sigma\, dZ_t - L\, dN_t,
    \end{equation}
    where $Z_t$ is a Brownian motion, $N_t$ is a Poisson process with intensity
    $\lambda$, and $L \in (0,1)$ is the proportional loss upon a hack. Here $\mu$
    denotes the \emph{conditional-no-jump excess return}. Under the reference
    intensity $\lambda$, the unconditional expected excess return is $\mu - \lambda L$.
\end{description}

Let $w_t$ denote the share of wealth allocated to the risky DeFi asset and let $C_t$
denote consumption. Then wealth evolves as
\begin{equation}
    \frac{dW_t}{W_{t-}} = \left(r + w_t \mu - \frac{C_t}{W_t}\right)dt + w_t \sigma dZ_t - w_t L dN_t.
\end{equation}

\subsection{Preferences and ambiguity}

We work with the time-separable CRRA special case of a robust-control problem. The
investor chooses $(C_t,w_t)$ while Nature chooses a distortion $\xi_t>0$ that scales
the jump intensity from $\lambda$ to $\lambda_t^*=\xi_t\lambda$, subject to an
entropy penalty. The stationary value function solves
\begin{equation}
    J(W) = \sup_{C,w}\inf_{\xi>0}\,
    \mathbb{E}^{\mathbb{Q}_\xi}\left[\int_0^\infty e^{-\delta t}
    \left(
        \frac{C_t^{1-\gamma}}{1-\gamma}
        + \Psi \lambda (\xi_t \ln \xi_t - \xi_t + 1)
    \right)dt
    \right].
\end{equation}
Here $\gamma>0$ is relative risk aversion and $\Psi>0$ is ambiguity tolerance, so
lower $\Psi$ means stronger ambiguity aversion.

\subsection{HJB equation}

The stationary HJB is
\begin{equation}
\begin{aligned}
0 = \sup_{C,w}\inf_{\xi>0}\Bigg\{
    \frac{C^{1-\gamma}}{1-\gamma}
    -\delta J(W)
    + J_W(W)\big[W(r+w\mu)-C\big]
    + \frac{1}{2}J_{WW}(W)W^2 w^2 \sigma^2 \\
    + \xi\lambda\big[J(W(1-wL)) - J(W)\big]
    + \Psi \lambda (\xi \ln \xi - \xi + 1)
\Bigg\}.
\end{aligned}
\end{equation}

This expression uses the \emph{uncompensated} Poisson representation consistently.
In particular, there is no additional drift term subtracting $\xi\lambda wL$;
expected jump losses are accounted for only through the jump operator.

\subsection{Worst-case intensity distortion}

The first-order condition for the inner minimization with respect to $\xi$ is
\begin{equation}
    \lambda\big[J(W(1-wL)) - J(W)\big] + \Psi \lambda \ln \xi = 0.
\end{equation}
Hence the worst-case distortion is
\begin{equation}
    \ln \xi^* = \frac{J(W) - J(W(1-wL))}{\Psi},
    \qquad
    \xi^* = \exp\left(\frac{J(W) - J(W(1-wL))}{\Psi}\right).
\end{equation}
Because hack losses reduce wealth and $J(\cdot)$ is increasing, we have
$J(W) > J(W(1-wL))$, so $\xi^*>1$. Ambiguity therefore raises the investor's
subjective jump intensity relative to the reference intensity.

\subsection{Portfolio choice}

Assume CRRA form
\begin{equation}
    J(W)=A\frac{W^{1-\gamma}}{1-\gamma},
\end{equation}
with $A>0$. Then
\begin{equation}
    \frac{J_W(W(1-wL))}{J_W(W)} = (1-wL)^{-\gamma}.
\end{equation}
Using the envelope theorem for the inner minimization, the first-order condition for
$w$ becomes
\begin{equation}
    0 = \mu - \gamma w \sigma^2 - \xi^* \lambda L (1-wL)^{-\gamma}.
\end{equation}

The last term is the ambiguity-adjusted marginal jump-risk cost. As ambiguity
aversion rises (lower $\Psi$), $\xi^*$ increases and the optimal risky weight falls.

\begin{proposition}[Ambiguity-driven flight-to-quality]
There exists a region of the parameter space such that sufficiently large ambiguity
aversion, jump severity, or perceived exploit intensity pushes the investor toward a
corner allocation with
\begin{equation}
    w_{\text{DeFi}}^* = 0, \qquad w_{\text{safe}}^* = 1.
\end{equation}
\end{proposition}

This proposition should be read qualitatively: the model rationalizes why exploit
shocks can generate discrete reallocation into safer assets when ambiguity becomes
sufficiently important.

\subsection{Mapping to the main-text parameter $\eta$}

The main text uses $\eta$ as a \emph{demand-amplification parameter}. In the
stylized panic region above, the gross demand response for the safer asset can be
approximated as increasing in the subjective jump intensity
$\lambda^*=\xi^*\lambda$. This motivates the approximation
\begin{equation}
    \eta \approx \xi^* > 1,
\end{equation}
not as an exact identity but as a crisis-region mapping: ambiguity can cause gross
demand for safer assets to exceed the direct physical redemption shock.

Two clarifications follow.

First, $\eta$ is not a substitution elasticity. It is a ratio-like amplification
object linking gross safe-asset demand to exploit-linked redemption pressure.

Second, if the normalized supply pressure in the CP segment is $R_t$, then gross
flight-to-quality demand is $\eta R_t$ and net excess demand under the normalization
used in the main text is $(\eta-1)R_t$.

\paragraph{Scope.}
This appendix motivates the sign and amplification logic used in the paper. It does
not, by itself, identify $\eta$ from the data. The empirical calibration of $\eta$
in the main text remains conditional on the assumed price-impact parameter
$\lambda_H$.

\section{A Stylized Global-Game Rationale for State Dependence}

This appendix provides a stylized rationale for why exploit-window redemptions may
be state dependent when congestion costs and coordination motives interact. The
objective is qualitative: to show why a nonlinear response may arise in principle.
The appendix does \emph{not} claim that the econometric cutpoint from the main text
is a structurally identified causal threshold.
The coordination logic follows the global-games tradition of \citep{morris1998unique,morris2003global}, adapted here to exploit-window redemptions and congestion costs.

\subsection{Environment}

Consider a continuum of investors $i\in[0,1]$, each deciding whether to redeem
immediately ($a_i=1$) or wait ($a_i=0$). Let $\theta$ denote the latent
post-exploit solvency or confidence state of the protocol, with lower $\theta$
indicating weaker fundamentals. Agent $i$ receives noisy private signal
\begin{equation}
    x_i = \theta + \sigma \varepsilon_i, \qquad \varepsilon_i \sim N(0,1).
\end{equation}
Let
\begin{equation}
    A = \int_0^1 a_i\,di
\end{equation}
denote aggregate redemption pressure.

\subsection{Payoffs}

Waiting yields
\begin{equation}
    u(0,A,\theta)=
    \begin{cases}
        1, & A<\theta,\\
        0, & A\ge \theta.
    \end{cases}
\end{equation}

Running avoids future insolvency risk but incurs congestion costs:
\begin{equation}
    u(1,A,\theta)=1-C(A),
\end{equation}
where
\begin{equation}
    C(A)=\phi_0+\gamma A^\lambda, \qquad \gamma>0,\ \lambda\ge 1.
\end{equation}
Higher aggregate redemptions raise congestion costs, so congestion acts as a
strategic substitute to the coordination motive in runs.

\subsection{Threshold intuition}

In the standard global-game limit with precise enough information, there is a unique
switching threshold $\theta^*$: investors redeem when their signal is sufficiently
pessimistic relative to that threshold. In reduced form, two forces shape
$\theta^*$:

\begin{itemize}
    \item \textbf{Congestion effect:} higher expected congestion costs make running
    less attractive and push the run threshold downward;
    \item \textbf{Ambiguity effect:} greater uncertainty about exploit severity makes
    waiting less attractive and pushes the run threshold upward.
\end{itemize}

These opposing forces imply that exploit-linked redemption pressure need not vary
linearly with congestion. In ordinary states, rising gas can dampen running by
making immediate exit costly. In more stressed states, ambiguity can dominate and
redemption pressure can remain high despite elevated gas.

\subsection{Connection to the empirical section}

This appendix motivates why the empirical data may display \emph{state dependence}
in exploit-window redemptions. It does not imply that the econometric cutpoint in
the exploratory threshold regression is a structural or causal threshold. The
main-text evidence on gas-state nonlinearity should therefore be interpreted
descriptively, as suggestive of state dependence rather than as a calibrated
global-game equilibrium.

\end{document}